\documentclass[aps,prd,amsmath,10pt,amssymb,twocolumn,showpacs,superscriptaddress]{revtex4-1}

\usepackage{hyperref}
\usepackage{graphicx}
\usepackage{amsfonts,amsmath,amssymb,bm}
\usepackage{array}
\usepackage{color}

\begin{document}


\title{Spallation backgrounds in Super-Kamiokande are made in muon-induced showers}

\author{Shirley Weishi Li}
\affiliation{Center for Cosmology and AstroParticle Physics (CCAPP), Ohio State University, Columbus, OH 43210}
\affiliation{Department of Physics, Ohio State University, Columbus, OH 43210}

\author{John F. Beacom}
\affiliation{Center for Cosmology and AstroParticle Physics (CCAPP), Ohio State University, Columbus, OH 43210}
\affiliation{Department of Physics, Ohio State University, Columbus, OH 43210}
\affiliation{Department of Astronomy, Ohio State University, Columbus, OH 43210 \\
{\tt li.1287@osu.edu,beacom.7@osu.edu} \smallskip}

\date{April 28, 2015}

\begin{abstract}

Crucial questions about solar and supernova neutrinos remain unanswered.  Super-Kamiokande has the exposure needed for progress, but detector backgrounds are a limiting factor.  A leading component is the beta decays of isotopes produced by cosmic-ray muons and their secondaries, which initiate nuclear spallation reactions.  Cuts of events after and surrounding muon tracks reduce this spallation decay background by $\simeq 90\%$ (at a cost of $\simeq 20\%$ deadtime), but its rate at 6--18 MeV is still dominant.  A better way to cut this background was suggested in a Super-Kamiokande paper [Bays {\it et al.}, Phys.~Rev.~D {\bf 85}, 052007 (2012)] on a search for the diffuse supernova neutrino background.  They found that spallation decays above 16 MeV were preceded near the same location by a peak in the apparent Cherenkov light profile from the muon; a more aggressive cut was applied to a limited section of the muon track, leading to decreased background without increased deadtime.  We put their empirical discovery on a firm theoretical foundation.  We show that almost all spallation decay isotopes are produced by muon-induced showers and that these showers are rare enough and energetic enough to be identifiable.  This is the first such demonstration for any detector.  We detail how the physics of showers explains the peak in the muon Cherenkov light profile and other Super-K observations.  Our results provide a physical basis for practical improvements in background rejection that will benefit multiple studies.  For solar neutrinos, in particular, it should be possible to dramatically reduce backgrounds at energies as low as 6 MeV.
\end{abstract}  


\maketitle


\section{Introduction}
\label{sec:intro}

Neutrino astronomy in the MeV range has been very successful.  Measurements of solar neutrinos confirmed many aspects of the nuclear fusion reactions that power the Sun; they also provided essential information about neutrino mass and mixing, especially the matter-induced effects.  The detection of neutrinos from SN 1987A and the identification of its progenitor star together confirmed the prediction that Type II supernovae arise from the collapse of the core of a massive star into a proto-neutron star; the extreme conditions allowed many novel tests of neutrino properties.

However, there are unresolved questions about the Sun and supernovae that can only be answered with improved sensitivity.  A better measurement of $^8$B neutrinos could improve knowledge of the solar core temperature, test the energy dependence of the electron-neutrino survival probability, and strengthen the signal of the day-night effect (presently 3 $\sigma$~\cite{Blennow2004,Renshaw2014})~\cite{GonzalezGarcia2008,Adelberger2011,Antonelli2012,Haxton2013}.  A first detection of the $hep$ flux would provide new tests of the solar model and neutrino mixing.  An eventual Milky Way supernova will allow high-statistics tests of the physical conditions attending neutron-star birth, flavor mixing in extreme conditions, and possibly black hole formation~\cite{Duan2010,Janka2012,Scholberg2012,Adams2013,Burrows2013}.  An immediate goal is the first detection of the diffuse supernova neutrino background (DSNB), which will provide new insights about supernova neutrino emission and the cosmic star formation history~\cite{Beacom2010}.

Discoveries could be made with existing experiments if detector backgrounds were reduced.  We focus on Super-Kamiokande (Super-K), by far the largest low-energy neutrino detector, with 22.5 kton of pure water in its fiducial volume (FV)~\cite{Fukuda2003,Abe2014}.  Great success in reducing backgrounds has already been achieved, but further gains have been stubborn.  For the robustly detected solar-neutrino signal, the signal/background ratio is only $\sim 0.1$ after standard cuts; at forward angles relative to the Sun, the ratio is $\sim 1$~\cite{Hosaka2006,Cravens2008,Abe2011}.  For the DSNB search, the high background rate means that the analysis energy threshold is above the peak energy of the signal spectrum~\cite{Malek2003,Bays2012,Zhang2015}.  Decreasing the background rate by a factor $\gtrsim 10$ would substantially advance solar neutrino studies and the DSNB search.  Is this possible without building a bigger, deeper detector?  Yes.

After standard cuts, the dominant background in the Super-K FV between 6--18 MeV is beta decays of nuclear spallation products~\cite{Koshio,Gando2003,Hosaka2006,Cravens2008,Abe2011}, which are short-lived isotopes produced from oxygen in association with cosmic-ray muons.  (At lower energies, longer-lived isotopes produced through radon ingress and decay are dominant.)  When a muon passes through Super-K, a cut around the measured position of the muon track is made to reject the spallation decays that follow; a difficulty is that some decay lifetimes are long (up to 20 s; see Table~I in Ref.~\cite{Li2014}) compared to the average time between muons ($\simeq 0.5$ s).  More precisely, a likelihood method is used to test events based on time elapsed since the muon, distance from the track, and a variable related to muon energy loss.  The empirical cut that Super-K has developed for solar neutrino studies effectively removes $\simeq 90\%$ of the backgrounds but introduces $\simeq 20\%$ deadtime, making it hard to improve.

In a previous paper~\cite{Li2014}, we performed the first theoretical calculation of the production and properties of the spallation decay backgrounds for water-based Cherenkov detectors such as Super-K.  Our predictions are in good agreement, within a factor of 2, with Super-K data on the energy spectrum and time profile for the sum of spallation decay isotopes, and could be improved by calibration and more careful comparison.  (Comparable accuracy is found in spallation studies for scintillator detectors~\cite{Abe2010,Bellini2013}.)  We detailed the physical processes behind isotope production and ways to use this knowledge to improve cuts.  An important point is that {\it nearly all isotopes are produced not by the muons themselves, but by the secondary particles associated with their energy-loss processes}.  At the depth of Super-K (2700 m water equivalent), where the average muon energy is 270 GeV, the average energy loss for a vertical throughgoing muon is 11 GeV, of which 7 GeV is from continuous processes such as ionization and 4 GeV from radiative processes such as delta-ray production and bremsstrahlung.  Fluctuations can make the radiative losses much larger.

A recent Super-K paper on the DSNB search~\cite{Bays2012} showed that the Cherenkov light yield associated with a muon varies along its track, exceeding that expected for a single muon and presenting a broad peak (comparable in length to the height of the FV), and that subsequent spallation decays are correlated in position with this peak.  The reasons for this variation, its properties, and its association with spallation decays went unexplained.  However, it was found that these facts could be exploited to improve the rejection of spallation decays.  Using an effectively shorter section of the muon track, several times less than the height of the FV, a more aggressive cut was used while keeping the deadtime moderate.  This allowed Super-K to lower the analysis threshold for the DSNB search from 18 to 16 MeV, with zero spallation events remaining.

Here we provide the first explanation of the physics behind the Super-K technique, as well as new insights to substantially improve its effectiveness.  Because the Cherenkov intensity (light emitted per unit length) of a relativistic muon is constant, the extra light and its variation must be due to additional charged particles, and a natural explanation is that these are produced in showers.  However, the variations shown by Super-K (Fig.~2 in Ref.~\cite{Bays2012}) and Fig.~4.2 in Ref.~\cite{Bays} appear to be grossly inconsistent with this explanation, because the spatial extent is too large and the amplitude too small.  Nevertheless, we find that the excess light is indeed due to particles in showers; that these showers are of short extent with high light intensity but appear long with low intensity due to Cherenkov reconstruction issues; that the correlation between the light profile peak and spallation production is because nearly all isotopes are made in showers; and that these reconstructions can be improved.  Using our results, Super-K could refine their new cut down to 6 MeV to improve solar neutrino and DSNB studies.

The framework for our calculations closely follows that of our previous paper, and details are given there~\cite{Li2014}.  We use the particle transport code FLUKA~\cite{Ferrari2005,Battistoni2007} (version 2011.2b.6) for our calculations, which has been used extensively for simulating muon-induced backgrounds in underground detectors~\cite{Wang2001,Kudryavtsev2003,Galbiati2005a,Galbiati2005,Mei2006,Abe2010,Empl2012,Empl2014,Bellini2013}.  We use the same physics choices for FLUKA, details of the Super-K geometry setup, and the muon spectrum.  Our calculations are for single throughgoing muons traveling 32.2 m vertically down the center of the FV.  We assume that the positions of muon tracks are always well determined by a combination of outer-detector and inner-detector information, aided by the long lever arm of the muon track.  For nonvertical throughgoing muons with shorter path lengths or for muon bundles, our results could be adjusted appropriately.  We discuss stopping muons separately.

Whereas our previous paper considered the average behavior of muons (from one to the next, and along each track), we now follow the energy-loss variations of individual muons.  We separately simulate how muons create daughter particles, how these daughters induce showers, and how these showers produce isotopes.  With our new approach, we recover our previous results.  All particles eventually produced following a muon are called secondaries; those in the first generation are called daughters.

The scope of this work is defined by a few choices.  We focus on explaining and extending the results of Ref.~\cite{Bays2012}.  We explain just the main features of the Super-K results; improving the details would require further input from them.  We do not yet attempt a full calculation of the reduction in backgrounds; our estimates are enough to show the promise of new techniques.  In our next paper, we will show why the Super-K Cherenkov reconstruction results appear to be inconsistent with showers and how they can be improved.

The remainder of this paper is organized as follows.  In Sec.~\ref{sec:showers}, we focus on the physics of showers --- the energy spectra of their secondaries, their geometric properties, and the rates of showers as a function of their energy --- to highlight physics insights critical to understanding later results.  In Sec.~\ref{sec:correlation}, we detail how isotopes are produced and how this explains the observed correlation between Cherenkov light yield and spallation decays.   Finally, we conclude in Sec.~\ref{sec:conclusion}.


\section{Shower Physics}
\label{sec:showers}

\begin{figure*}[th]
    \begin{center}                  
        \includegraphics[width=\textwidth]{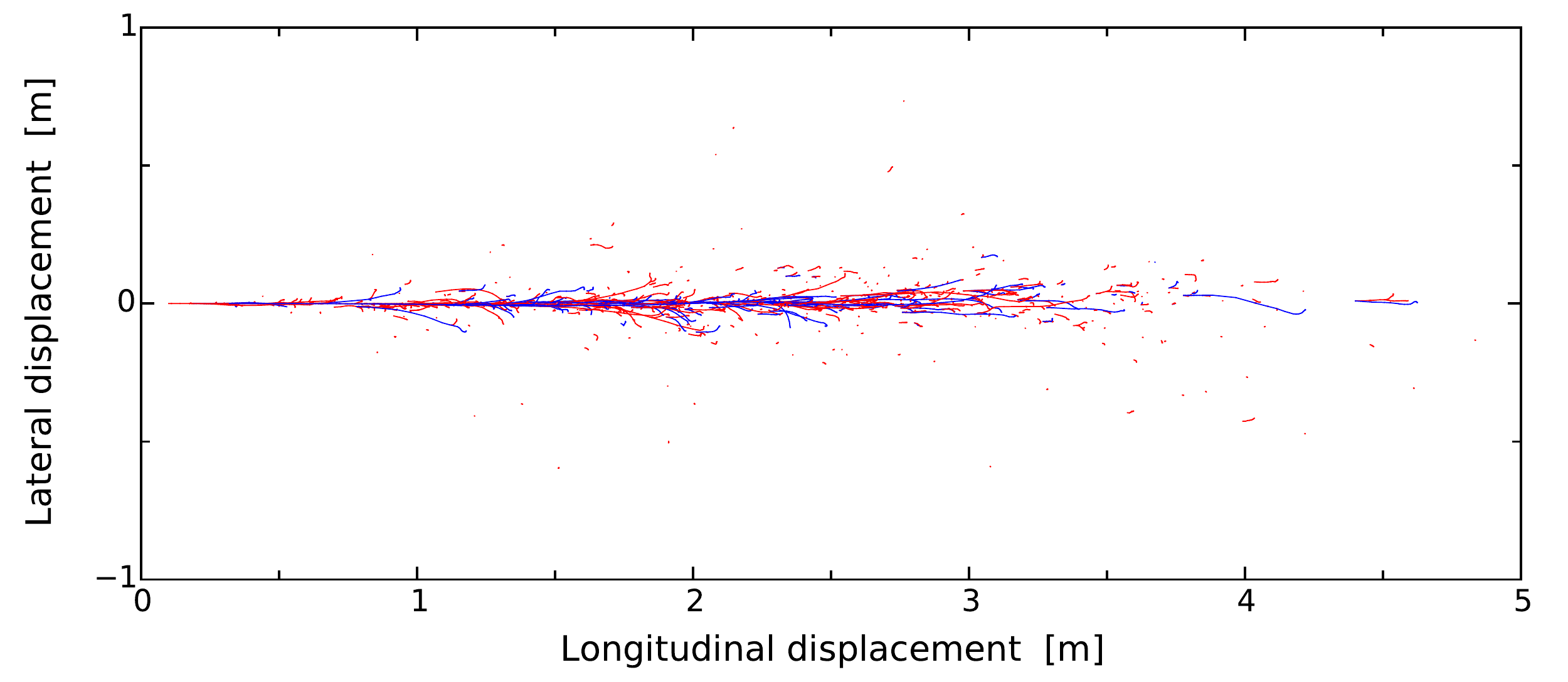}
        \caption{An electromagnetic shower in water initiated by a 10 GeV electron.  The red lines are electrons, and the blue lines are positrons.  The $x$ and $y$ axis ranges are chosen to not distort the relative lateral and longitudinal scales.}
        \label{real_shower}
    \end{center}
\end{figure*}

Particle shower (or cascade) processes are central to this paper.  Showers can be produced by radiative energy losses of cosmic-ray muons in Super-K, especially at high energies.  The basic physics is that particles multiply in number through repeated interactions, with the particle energy decreasing in each generation.  This continues until the average energy drops below a critical energy $E_c$ that depends on the type of shower and the medium.  Below this, charged particles mostly lose energy by ionization.

For electromagnetic showers, the main secondary particles are electrons, positrons, and gamma rays.  The dominant interactions in water are electrons and positrons producing gamma rays through bremsstrahlung with nuclei, and gamma rays pair-producing electrons and positrons, also with nuclei.

For hadronic showers, the main secondary particles are charged and neutral pions.  Hadron interactions with nucleons produce pions, and pion interactions with nucleons can change both of their charges.  The basic processes in electromagnetic showers leave the target nuclei largely intact, but that is not true for hadronic showers, which brings additional complications.

Figure~\ref{real_shower} shows a typical shower.  Shower lengths are around a few meters; shower widths are around tens of centimeters.  Most electrons and positrons in showers are forward, with $\langle\cos\theta_z\rangle \simeq$ 0.9.

Showers are defined most generally by the phase-space density of their secondary particles, i.e., the joint number density in momentum and position, with time as a parameter.  To express the cumulative effects of a shower, integrated over time, we use not the number density of secondary particles, which is only defined at a given instant, but rather some measure of their integrated effects.  For a Cherenkov detector, it is useful to weight by path length; for charged particles, this is proportional to the light produced (and, especially for electrons, is nearly proportional to the energy deposited).  Different integrals of the phase-space density are convenient for different purposes.  The path length profile in longitudinal position (integrating over momenta and lateral positions) is probably the most familiar, and it determines the observable muon light profile in Super-K.  The path length spectrum in energy (integrating over positions and the momentum directions) is not commonly shown, but it determines isotope production in Super-K.  We present these in the opposite order, covering path length spectra in Sec.~\ref{subsec:em} and~\ref{subsec:hadronic} and longitudinal profiles in Sec.~\ref{subsec:geometry}.

To provide more detail on path length spectra, $\mathrm{d}L/\mathrm{d}E$ {\it describes the sum of distances traveled by all particles of a given species at each energy}.  This is obtained by integrating over the positions of the particles, and is called the volume-integrated fluence in FLUKA~\cite{Papiez1994,Ferrari2005,Li2014}.  This spectrum multiplied by the cross section as a function of energy is the integrand for calculating the interaction rate.  The integrated path length above the Cherenkov threshold determines the total Cherenkov intensity and thus the number of photomultiplier tube hits.

Super-K, a water-based Cherenkov detector, directly observes only relativistic charged particles.  We focus on the light produced by showers induced by muons.  The muons themselves produce Cherenkov light at constant intensity along the muon track~\cite{Jackson}.  Super-K cannot separate electrons from positrons or $\pi^-$ from $\pi^+$, so, hereafter, electrons means the sum of electrons and positrons and pions means the sum of $\pi^-$ and $\pi^+$, unless specified otherwise.  Charged particles below their Cherenkov thresholds (kinetic energy 0.257 MeV for electrons and 70.1 MeV for pions~\cite{Koshio}) are not detectable.  Gamma rays and neutrons are not detectable directly, but only through their interactions.

We do not discuss isotope production by showers in this section.  However, it is helpful to keep in mind that the most important parent particles for background isotopes are neutrons and pions; gamma rays make a small fraction of isotopes and electrons do not make isotopes~\cite{Li2014}.  Hence, even though neutrons, pions, and gamma rays contribute negligibly to Cherenkov light production, we discuss their behavior in showers.

In the remainder of this section, we first study the physics of showers in water independent of primary muons.  Then we discuss how cosmic-ray muons make daughter particles and thus showers with a variety of energies in Super-K.


\subsection{Electromagnetic shower spectra}\label{subsec:em}

Some important aspects of electromagnetic showers can be understood using simple principles.  In a model proposed by Heitler~\cite{Heitler1954}, it is assumed that bremsstrahlung and pair production have the same mean free path (radiation length $X_0$), that this is energy independent, and that all other interactions, including electron ionization, can be ignored.  Further, it is assumed that in each generation, particles travel the same fixed distance ($d = X_0\ln{2}$) before they split into two particles, each with half the parent particle energy.  

\begin{figure}[t]
    \begin{center}                  
        \includegraphics[width=\columnwidth]{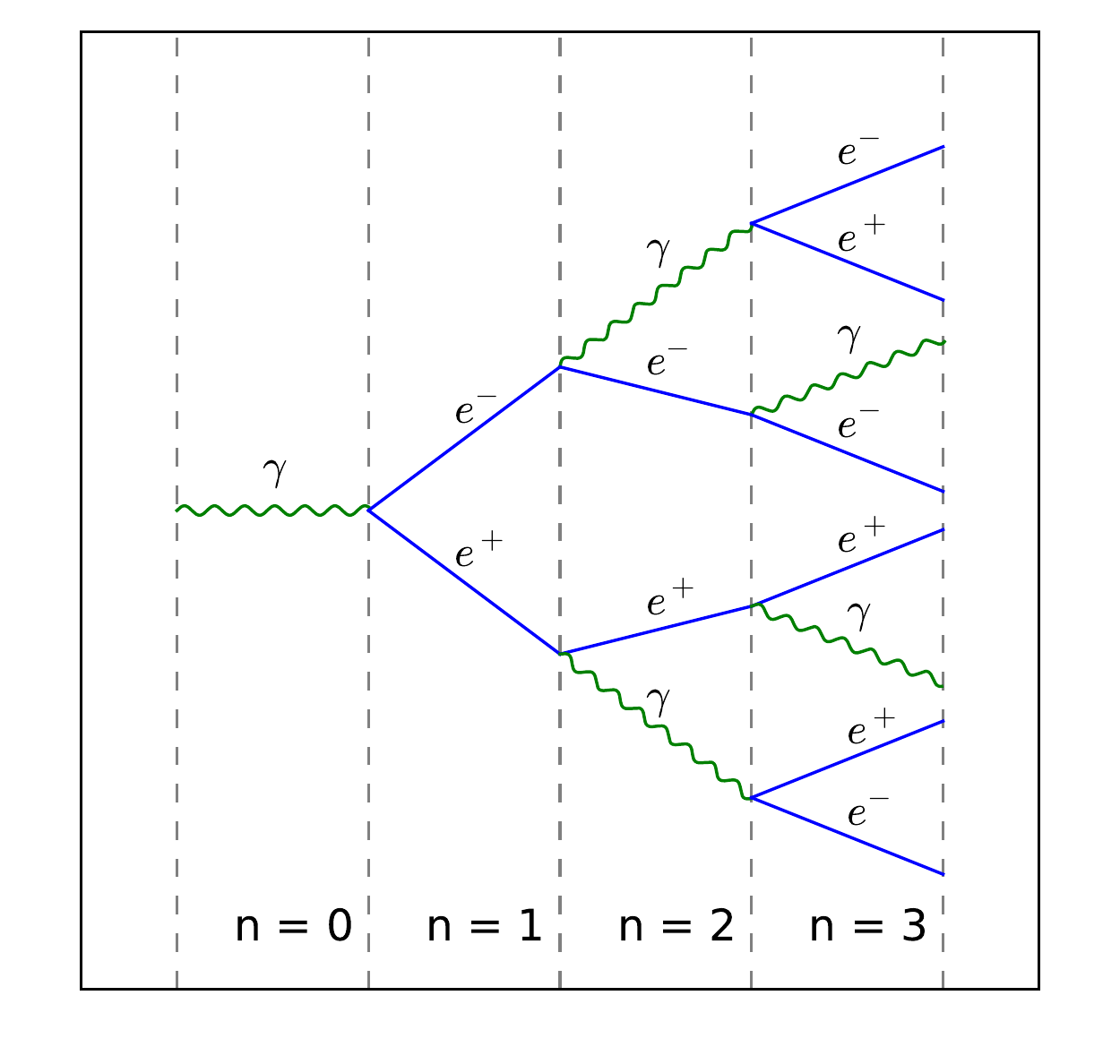}
        \caption{Schematic diagram of Heitler's model for electromagnetic showers in the growing phase, which continues until the particle energies are below $E_c$.}
        \label{em_shower}
    \end{center}
\end{figure}

Figure~\ref{em_shower} illustrates this process.  If the shower starts with one particle of energy $E_0$, then after $n$ generations, there are $2^n$ secondary particles, each with energy 
\begin{equation}
	E_n = \frac{E_0}{2^n}\,.
\end{equation}
The shower stops growing when the average particle energy is below the critical energy $E_c$, which is set by the electron ionization energy loss in one radiation length~\cite{pdg}.  Then, a shower reaches its maximum, where the number of particles is the greatest, after $\log_2(E_0/E_c)$ generations.  Because the particles are mostly forward due to being relativistic, the distance to the shower maximum is
\begin{equation}\label{Heitler}
	\ell = d \log_2\!\left(\frac{E_0}{E_c}\right) = X_0 \ln\!\left(\frac{E_0}{E_c}\right).
\end{equation}
In water, $X_0 = 36$ cm and $E_c = 80$ MeV~\cite{pdg}.  Electrons with energy $E_c$ lose all of their energy by ionization in one radiation length.  After shower maximum, gamma rays and the electrons they scatter will travel somewhat further (a few radiation lengths).  For a 10 GeV shower, the longitudinal extent of a shower would be $\sim$ 2 m, far less than the height of Super-K.  The true shower extent is greater than this, but not much, and is discussed in Sec.~\ref{subsec:geometry}.

Further properties of showers can be obtained analytically with more complex models~\cite{Carlson1937,Bhabha1937,Rossi1941,Greisen1956,Greisen1960,Lipari2009,Lipari2009b}.  An example of the latter is the work by Rossi and Greisen~\cite{Rossi1941}, where they derived results by solving the Boltzmann equations under certain assumptions.  In their Approximation A, which is only valid for high particle energies, asymptotic cross sections for bremsstrahlung and pair production are assumed and electron ionization energy loss is neglected.  For the electron path length spectrum in an electron-initiated shower, they find
\begin{equation}
	\frac{\mathrm{d}L}{\mathrm{d}E} = 0.437 X_0\,\frac{E_0}{E^2}
\end{equation}
for electron energies $E \gg E_c$.  For electrons with energy greater than $E$, the distance to their maximum is
\begin{equation}\label{showermax}
	\ell = 1.01 X_0 \left(\ln\!\left(\frac{E_0}{E}\right)-1\right) .
\end{equation}
This is similar to the Heitler result if Eq.~(\ref{showermax}) is (inappropriately) evaluated at $E_c$.

\begin{figure}[t]
    \begin{center}                  
        \includegraphics[width=\columnwidth]{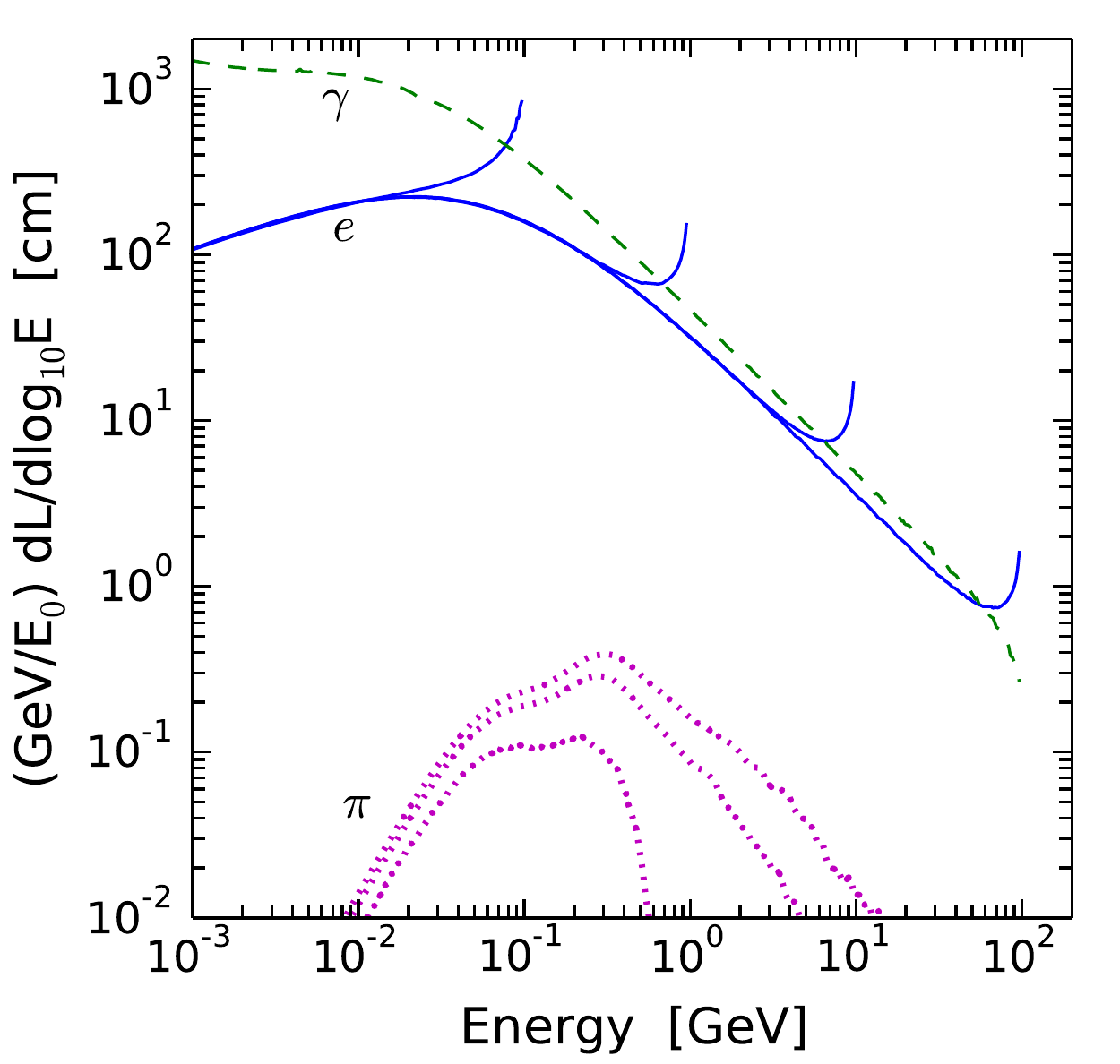}
        \caption{Electron, gamma ray and pion path length spectra in terms of kinetic energy for showers initiated by electrons of energy $E_0 =$ 0.1, 1, 10, and 100 GeV.  The features seen at the injection energy arise because the showers have not yet reached an equilibrium mixture of $e^-$, $e^+$, and $\gamma$.  The gamma-ray path length is shown only for $E_0 = 100$ GeV; the other cases are similar, except for having lower endpoints.  The pion path length spectra are  shown for $E_0 =$ 1, 10, and 100 GeV (it is zero for 0.1 GeV).  All spectra are normalized by $E_0$.}
        \label{shower_comparison}
    \end{center}
\end{figure}

Contemporary work on showers is based on Monte Carlo simulation of all microscopic processes~\cite{Hillas1982,Zas1992,Razzaque2002,Akchurin2005}.  The fluctuations (distance, energy, etc.) in every interaction are taken into account, instead of solving for the average behavior with the Boltzmann equation.  This enables the study of individual showers, as well as the variations among them.  The simulation results are valid for the entire energy range, and the precision is excellent.  In the following, we use theoretical insights to illustrate the physics behind our numerical results.

Figure~\ref{shower_comparison} shows particle path length spectra for electron-initiated showers.  We inject electrons with fixed energies into the Super-K FV, which is large enough to contain all secondary particles.  We discuss Fig.~\ref{shower_comparison} from high to low energy.  As individual showers develop, the average energy of the shower particles decreases.  At the peak, which is somewhat below $E_c$, the particle number is at a maximum.  At lower energies, particles stop multiplying and the path length decreases due to particle ionization losses.  

{\it The way these and other results are shown is designed to highlight key physics points}.  As discussed, the numerator is the total path length traveled by a group of particles, and not just the number of particles.  We divide by the injection energy $E_0$ to show when there is universality (more energetic showers being just multiples of less energetic showers) or deviations from that.  Because of the large range of energies, we use a $\log$ scale on the $x$ axis; also, this is especially appropriate for the showering phase, where particle energies change by factors, not shifts, between each generation.  To calculate integrals of the curves, one should use $\log_{10}E$ as the integration variable.  To match this choice of axis, we take derivatives with respect to $\log_{10}E$, which makes the height of the curve proportional to its importance in the integral; note that $\mathrm{d}L/\mathrm{d}\!\log_{10} E = 2.3 E\mathrm{d}L/\mathrm{d}E$ (see Ref.~\cite{Beacom2010b,Li2014} for further discussion).  All energies in logarithms are in GeV units.  A $\log$ scale is often used on the $y$ axis.  This is of no particular importance, except that one should judge the relative contributions to the integral by numerical, not visual, height.

The spectra at high energies, during the shower phase, go as $\mathrm{d}L/\mathrm{d}\!\log_{10} E \sim 1/E$ for both electrons and gamma rays.  The differential cross sections for bremsstrahlung and pair production can be factorized to roughly depend only on the fractional energy of the outgoing particles~\cite{pdg}.  The path length spectra should be a function of $E/E_0$, and a power law shows this scale invariance~\cite{Lipari2009}.  The shower is extensive in (proportional to) $E_0$, so the length must be proportional to $E_0$.  The result must also scale linearly with the radiation length $X_0$.  Then, using simple dimensional analysis, we know the path length spectrum must scale as $\sim X_0 E_0/E^2$.  This is consistent with the results of Rossi and Greisen~\cite{Rossi1941}.  The slight difference between the gamma-ray and electron path lengths at high energies in Fig.~\ref{shower_comparison} is due to electron ionization, which matters more as the energy decreases.

The electron path length spectra at low energies, during the ionization phase, go as $\sim E^{0.5}$.  To first order, ionization conserves particle number, but dissipates energy in the shower, so we might expect $\mathrm{d}L/\mathrm{d}E \sim$ constant and $\mathrm{d}L/\mathrm{d}\!\log_{10} E \sim E$.  However, below the peak, there are many gamma rays from bremsstrahlung, as shown in Fig.~\ref{shower_comparison}, and these inject energy to electrons from the medium through Compton scattering.  The competition between this and ionization produces the electron spectrum shown, including shifting the peak to an energy below $E_c$.

For an injection energy of 0.1 GeV or lower, showers do not typically develop.  Electrons range out by ionization and do not produce or accelerate other particles.  Gamma rays undergo Compton scattering and pair production, but they do not produce particles other than electrons.

The hadronic particle content in electromagnetic showers is quite small on average, and the pion path lengths are a few orders of magnitude less than those for electrons.  The shapes of the pion spectra reflect the large pion mass and the large energy required for pion production by photo-nuclear interactions.  We discuss this in the next subsection.  

The electron path length spectra are nearly extensive in $E_0$ (same for the gamma-ray path lengths).  These lie on top of each other when we divide out this initial energy.  In other words, particles in an electromagnetic showers quickly lose information about the initial energy, and such showers are self-similar except for total energy~\cite{Lipari2009b}.  (This is less true for the hadronic components of the showers.)  Consequently, the total path lengths are extensive in $E_0$.  Because electron ionization is the dominant dissipative energy-loss process, the total path length of electrons in water is 
\begin{equation}
	L \simeq \frac{E_0}{2\,\mathrm{MeV/cm}} \,.
\end{equation}

For electromagnetic showers of fixed energy, the total path length for electrons does not fluctuate much.  For example, for a 10 GeV electron initiated shower, the average total electron path is $\simeq$ 5500 cm, while the standard deviation is only $\simeq$ 200 cm.  Most of the fluctuations arise from the rare production of hadronic components, for which there is some energy loss without Cherenkov light (e.g., neutrons, nonrelativistic protons).  In addition, there is some contribution to the fluctuations because the electron ionization rate depends on energy.

The Cherenkov light intensity is proportional to the electron path length.  Figure~\ref{shower_comparison} shows that most of the Cherenkov light comes from electrons near the critical energy~\cite{Lipari2009,Lipari2009b}.  The electron path length differences near the endpoints for different injection energies contribute negligibly to the total path length.  Also, there is little electron path length accumulated below the Cherenkov threshold.  Pion path lengths contribute negligibly because they are much shorter and pions have a higher Cherenkov threshold.  In sum, {\it the injection energy of an electromagnetic shower is accurately revealed by its total Cherenkov light}.  The visible energy of each shower is within a few percent of the true shower energy.

For gamma-ray-initiated showers, the path length spectra of particles (including the hadronic component) are almost identical to those of electron-initiated showers, except near the endpoint, because showers quickly lose information about the initial particle~\cite{Lipari2009b}.

Spallation isotopes are dominantly produced by particles that produce little (pions) or no (neutrons, gamma rays) Cherenkov light themselves.  However, these particles are accompanied by electrons through shower processes.  In Sec.~\ref{sec:correlation}, we detail how to exploit this connection and identify spallation products using Cherenkov light.


\subsection{Hadronic shower spectra}\label{subsec:hadronic}

In hadronic interactions in the GeV range and above, the dominant particles produced are pions, with roughly equal numbers of each charge.  Hadronic showers of $\pi^-$, $\pi^+$, and $\pi^0$ have much in common with electromagnetic showers of $e^-$, $e^+$, and $\gamma$, because both arise from particle multiplication processes and because the interaction lengths happen to be comparable.  Hadronic showers have a critical energy of about 1 GeV, where the probabilities for pions to multiply or to lose energy by ionization are equal.  The multiplicity of pions in hadronic showers increases with energy, being a few in the GeV range and a few tens in the TeV range~\cite{pdg}.  For further discussion, see Refs.~\cite{Giller2004,Matthews2005,Nerling2006,Engel2011,Montanus2014}, though note that their focus is on high energies and low densities.  In the following, we emphasize some differences between hadronic and electromagnetic showers.

{\it Although electromagnetic showers have, on average, only a small hadronic component, hadronic showers always have a dominant electromagnetic component}.  Charged pions interact, producing more pions and continuing the hadronic shower.  However, neutral pions promptly decay to gamma rays, feeding an electromagnetic shower.  With each new generation in the hadronic shower, roughly 1/3 of the remaining energy is transferred to the electromagnetic shower.  In principle, a hadronic shower with enough interactions would transfer all of its energy to the electromagnetic shower; in practice, the final hadronic fraction asymptotes at $\simeq 10\%$~\cite{Beatty2009,Rott2013}.  The number of charged pions reaching low energies is larger than would be naively expected due to large pion multiplicities at high energy and fluctuations in the energy division in each interaction.

\begin{figure}[t]
    \begin{center}
        \includegraphics[width=\columnwidth]{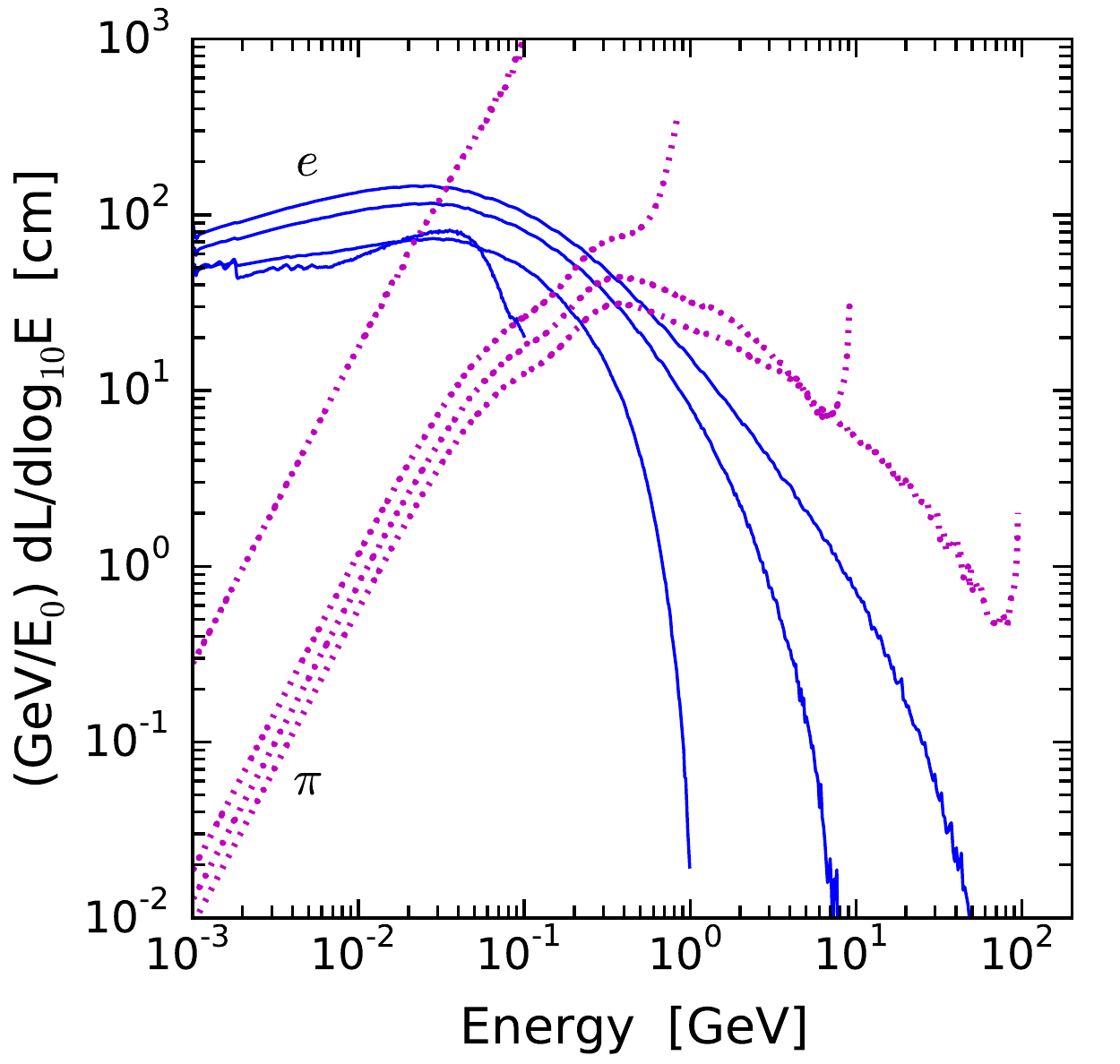}
        \caption{Electron and pion path length spectra in terms of kinetic energy for showers initiated by charged pions of energy $E_0 =$ 0.1, 1, 10, and 100 GeV.  Again, the features at the endpoints are injection effects.  The small features in the electron line for $E_0 =$ 0.1 GeV arise due to gamma rays from $\pi^0$ and nuclear decays.  All spectra are normalized by $E_0$.}
        \label{shower_comparison_pi}
    \end{center}
\end{figure}

Figure~\ref{shower_comparison_pi} shows particle path length spectra for showers initiated by charged pions.  For $E_0 =$ 1, 10, and 100 GeV, the primary pions have enough energy to induce hadronic showers; for the 0.1 GeV case, there is no pion multiplication and we discuss it separately.

The electron and pion spectra are not quite extensive in the injection energy.  This can be seen from the fact that the curves shown in Fig.~\ref{shower_comparison_pi} do not overlap.  With increasing injection energy, the fraction transferred to the electromagnetic shower increases.  For $E_0 =$ 1, 10, and 100 GeV pion-initiated showers, the fractional energy in electromagnetic showers is 31\%, 49\%, and 65\%.  The rest of the energy is dissipated by hadron and muon ionization energy loss, with a small fraction carried away by neutrinos.  Accordingly, as the injection energy increases, the pion curves fall and the electron curves rise.

Pion-initiated showers thus appear to be less energetic than electromagnetic showers with the same initial energy.  The visible energy is proportional to the total particle path length above the Cherenkov thresholds.  The energy that goes into the electromagnetic component of the shower produces Cherenkov light due to the $\simeq$ 500 cm / GeV of relativistic electron path length.  However, the energy that remains in the hadronic component of the shower is less efficient, with only $\simeq$ 100--200 cm / GeV of relativistic pion path length.  The difference is because some energy is lost to neutral particles and because pions become nonrelativistic at a higher energy than electrons.  In terms of light yield, pions are subdominant even in pion-initiated showers~\cite{Lipari2009}.  The visible energies for $E_0 =$ 1, 10 and 100 GeV pion showers are 0.57, 6.3, and 74 GeV.

The general features of the pion spectrum follow from the same principles that govern the electron spectrum: showering processes dominate at high energies, causing the increase in path length with decreasing energy, while ionization dominates at low energies, causing the decrease in path length with decreasing energy.  The critical energy for hadronic showers is higher than that for electromagnetic showers, due to the large pion mass and other factors, and the behavior of the path length spectrum in the peak region is more complex.  The peak near 0.4 GeV corresponds the most probable pion production energy.  At slightly lower energies, 0.1--0.3 GeV, some pions disappear through inelastic interactions of the form $\pi^- + p \rightarrow n$ and $\pi^+ + n \rightarrow p$ with bound nucleons, with the residual energy and momentum absorbed by their nuclei.  Once charged pions become nonrelativistic, the ionization rate increases quickly and the path length accumulated is small and decreases more steeply than for electrons below the peak.

When the pion injection energy is too low to create new pions, an electromagnetic shower cannot typically develop.  The pion path length spectrum is large, as all the energy remains with the pions, and this is the same for both $\pi^+$ and $\pi^-$.  For the $E_0 =$ 0.1 GeV case shown in Fig.~\ref{shower_comparison_pi}, the total pion path length is 23 cm.  Although this curve is much higher than the others, its integral is only slightly larger, corresponding to 230 cm / GeV, because nonrelativistic particles lose energy rapidly.  Rarely, a charged pion interacts with a nucleon and converts to a neutral pion, leading to some electromagnetic activity (on average 11 cm of electron path length).  Low energy $\pi^-$ are especially efficient at making isotopes through atomic and then nuclear capture~\cite{Ponomarev1973}; low energy $\pi^+$ do not efficiently make isotopes because they decay, not capture, once at rest.


\subsection{Shower geometry}\label{subsec:geometry}

\begin{figure}[t]
    \begin{center}                  
        \includegraphics[width=\columnwidth]{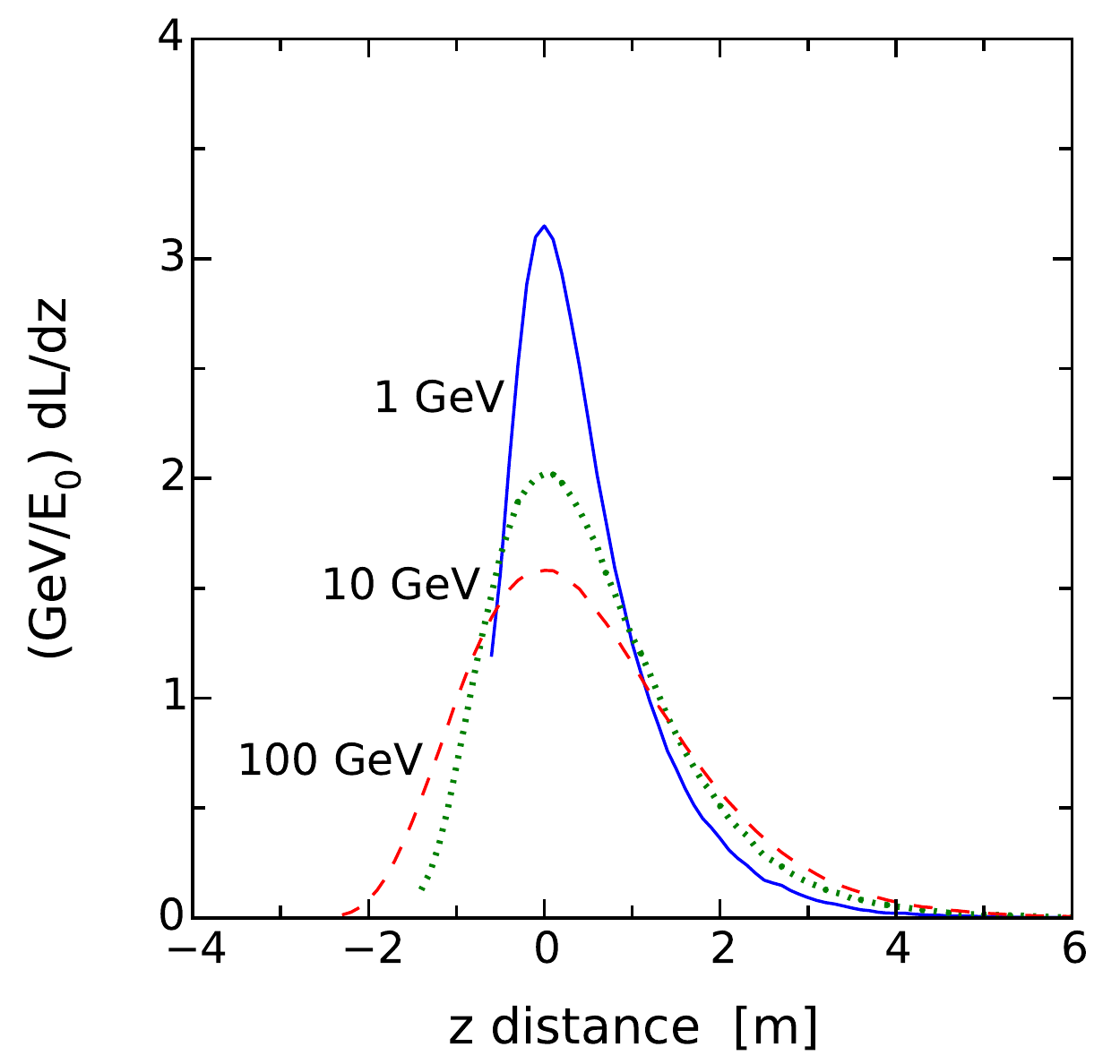}
        \caption{Average longitudinal profiles for showers initiated by electrons of energy $E_0 =$ 1, 10 and 100 GeV.  Here d$L$ is the charged-particle path length in all directions accumulated in a step d$z =$ 10 cm along the initial direction.  We separately shift the starting positions of the showers, each with one electron and height $\sim$ 1/$E_0$, so that the peaks line up at $z = 0$.  All profiles are normalized by $E_0$.}
        \label{shower_size}
    \end{center}
\end{figure}

\begin{figure*}[th!]
\begin{center}
    \includegraphics[width=\textwidth]{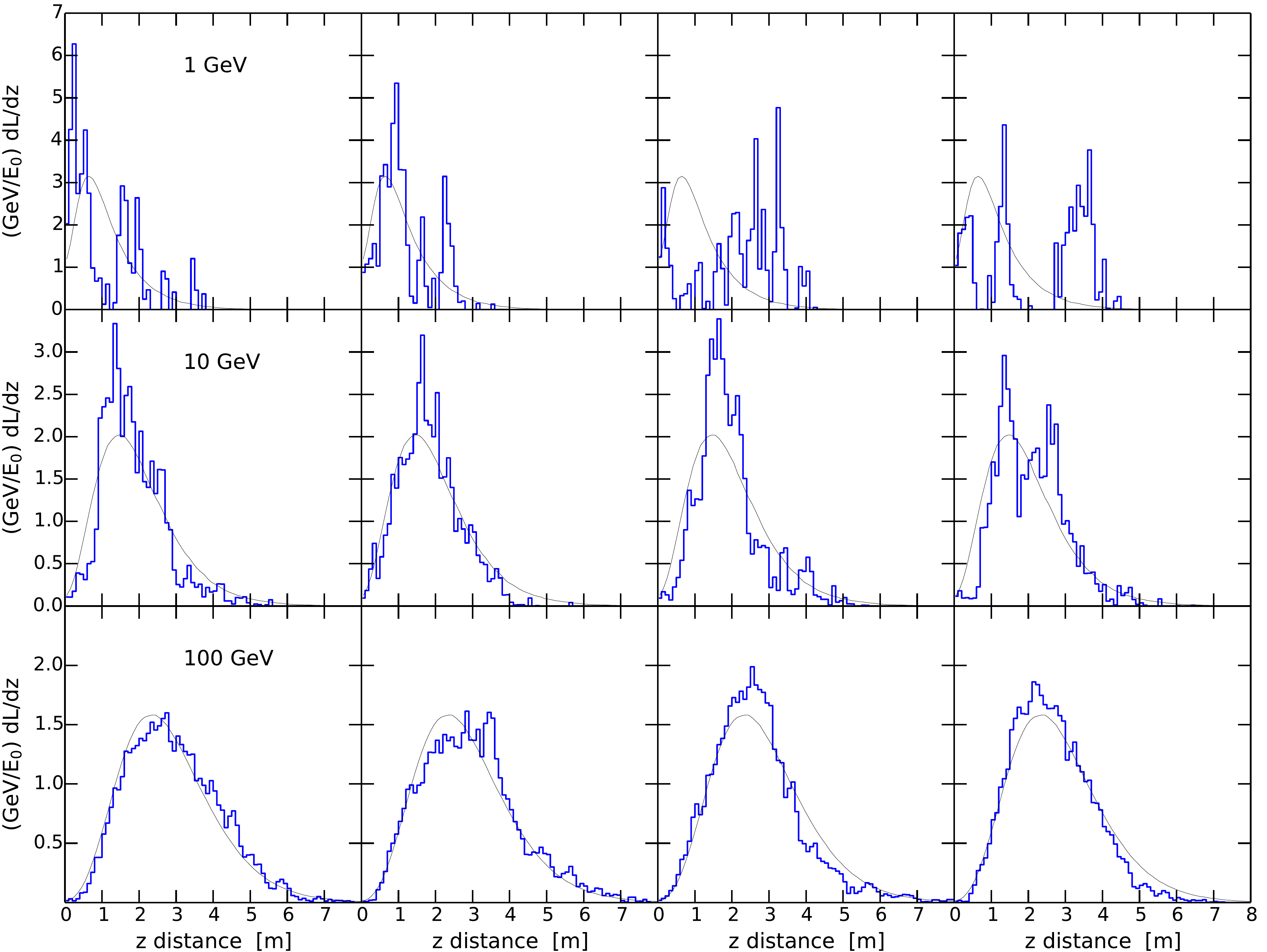}
    \caption{Examples of longitudinal profiles (blue bins) for showers initiated by electrons of energy $E_0 =$ 1, 10 and 100 GeV, as well as the averages (thin black lines).  All profiles are normalized by $E_0$.}
   \label{shower_example}
\end{center}
\end{figure*}

The physical distributions of showers and how they compare to the size of the Super-K detector are crucial for understanding why the new Super-K cut technique~\cite{Bays2012} gives such a big improvement.  The longitudinal and lateral sizes of showers define the region around the muon track where isotopes are made.  The exact profile and the deflection of shower particles determine the pattern of Cherenkov light.  We focus on electromagnetic showers in this section, because they are more common, because hadronic showers have a large electromagnetic shower component, and because hadronic showers are similar to electromagnetic showers in geometry (slightly different, and discussed below).

Figure~\ref{shower_size} shows the average longitudinal shower profile for three different injection energies.  We plot the electron path length per unit length along the initial direction, i.e., the Cherenkov intensity from the shower relative to that from a single particle.  This is roughly the instantaneous number of charged particles in the shower times (GeV/$E_0$).  This is not exactly true due to nonforward motion and particles starting or stopping within bins; in addition, these curves represent averages over many showers.  The area under the curve is the total electron path length scaled by the injection energy, and is nearly the same for all energies.  The showers extend 4--6 m for energies between 1--100 GeV.  This length is much shorter than the height of the Super-K FV, even for high-energy showers, which are rare.  

These average profiles show a rising phase, a peak, and a declining phase.  The distance to the peak position of the shower is an important parameter.  Even though Eq.~(\ref{Heitler}) and Eq.~(\ref{showermax}) were derived from simplified models, they are in good agreement with the full numerical results.  In more detail, the shape is consistent with standard formulas for the longitudinal profiles of showers, such as the Greisen~\cite{Greisen1956} and Gaisser-Hillas profiles~\cite{Gaisser1977}.

The overall profile shape, especially the length asymmetry between the rising and falling parts of the shower, is important for our discussions of shower correlations with spallation backgrounds in Super-K.  Compared to the naive Heitler model, where all electrons stop in one radiation length after shower maximum, the tails of realistic showers are long.  This arises from two types of fluctuations in showers: the distances particles travel before splitting obey an exponential distribution, and secondary particles do not always split the energy equally~\cite{Montanus2012}.  These fluctuations give a distribution to the particle energies in the shower at a given depth, instead of all particles having the same energy at the same location.  After the shower maximum, there are particles in the shower with energy higher than $E_c$ because they have interacted for fewer generations or because they have taken more energy from their parent particles.  These higher-energy particles stay in the shower longer, creating the long tail.

Figure~\ref{shower_example} shows examples of longitudinal profiles of individual showers, as fluctuations will affect shower reconstruction.  Showers with primary energies of 1 GeV look very different from one another and from the average profile.  With increasing initial energy, the relative fluctuations in shower profiles decrease.  Showers with 100 GeV have little variation in widths, peak position, and shape.  Because the shower energy is proportional to the Cherenkov light intensity, it is easy to measure the total energy in a shower (up to the ambiguity of whether it is electromagnetic or hadronic).  For high-energy showers, it might be possible to reconstruct them using the average profile as a template.  For low-energy showers, which are the most common, it is not clear if template fits will be helpful, due to the large fluctuations.

So far, we have simplified showers to be one dimensional and collinear.  Particles in showers do have lateral displacements.  The most important reason is electron displacement due to multiple scattering during propagation~\cite{Rossi1941}.  This is characterized by the Moli\`{e}re radius, which is about 10 cm in water~\cite{pdg}.  This is very small compared to either the Super-K muon track resolution or the distance between the spallation decay and the muon track.  The effects of the lateral extent of showers are negligible, so we skip discussions of their average profile or fluctuations.

However, though the lateral displacement of electrons is small on average, their angular deflections greatly affect how the shower appears in the detector.  Note from Fig.~\ref{real_shower} that individual electron paths are short but that deviations away from the forward direction are common.  We will discuss this in detail in our next paper.

As noted, hadronic showers are similar to electromagnetic showers in geometry, but there are some differences.  For 1 GeV hadronic showers, the longitudinal extent is similar to that shown in Fig.~\ref{shower_size}, but the shape is quite different.  Because this is so close to the hadronic critical energy, there are few generations, and we mostly see the average number of pions decrease according to an exponential set by the hadronic interaction length.  This might provide a way to identify low-energy hadronic showers, which are especially important for isotope production.  The longitudinal profiles for 10 and 100 GeV hadronic showers are quite similar to those of electromagnetic showers.  At all energies, the fluctuations in the longitudinal profiles of individual hadronic showers around the average are greater than for electromagnetic shower of the same energy; this might be used to distinguish hadronic showers on a statistical basis.  A more promising means might be to use the fact that hadronic showers have larger lateral extent (see Fig.~4 of Ref.~\cite{Li2014}).


\subsection{Shower frequency}

\begin{figure}[t]
    \begin{center}                  
        \includegraphics[width=\columnwidth]{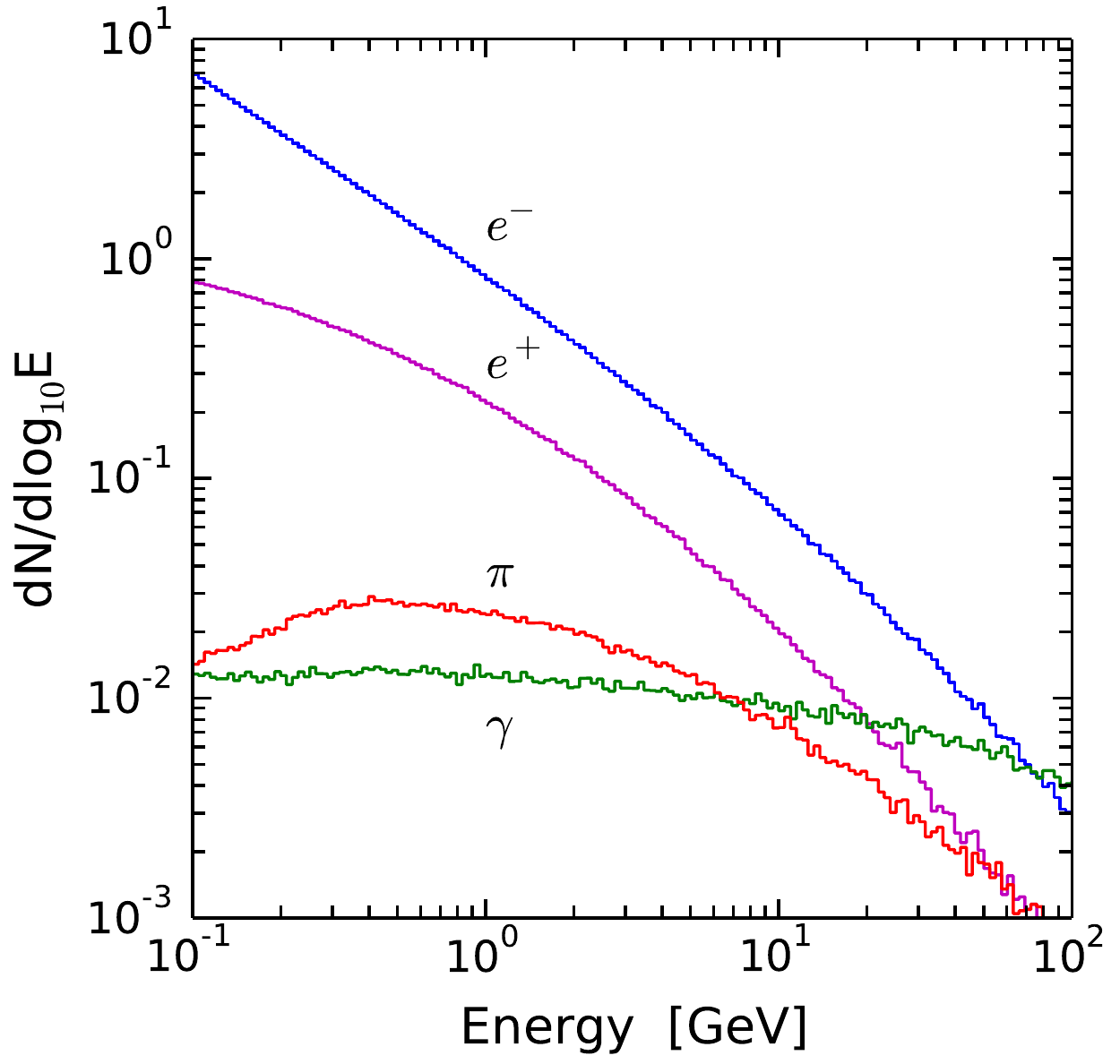}
        \caption{Daughter particle (first-generation secondary particle) kinetic energy spectra of electrons, positrons, gamma rays, and pions made directly by muons, normalized per muon, after convolution with the Super-K muon spectrum.  The $\pi$ line is the sum of $\pi^+$ and $\pi^-$; the $\pi^0$ line is about half of the $\pi$ line and is not shown.}
        \label{shower_spectrum}
    \end{center}
\end{figure}

Cosmic-ray muons abundantly produce daughter particles that initiate electromagnetic and hadronic showers.  Figure~\ref{shower_spectrum} shows the daughter particle production spectra obtained using the Super-K muon spectrum.  The frequencies are scaled by the muon rate in Super-K, and are thus numbers per muon.

The electron spectrum goes as $\mathrm{d}N/\mathrm{d}\!\log_{10} E \sim 1/E$.  This comes mainly from delta-ray production --- collisions of muons with atomic electrons where the energy transfer is large.  (Far more frequently, these collisions transfer little energy, and are treated as continuous ionization.)  For a muon energy of 270 GeV, the average at Super-K, the maximum energy transfer to an electron is 260 GeV~\cite{pdg}.  The differential cross section for delta-ray production scales as $\sim 1/E^2$ for electron energy transfers well below the maximum~\cite{pdg}.  This, plus the fact that we plot $\mathrm{d}N/\mathrm{d}\!\log_{10} E \sim E \mathrm{d}N/\mathrm{d}E$, largely explains the results shown.

The positron spectrum comes entirely from pair production, mostly through muon interactions with nuclei.  The differential cross section does not have a simple power-law form.  Using an approximate formula~\cite{Tannenbaum1991,Bulmahn2010}, we find that the differential cross section can be approximated by a broken power law: $\sim E^{-1.5}$ at low energies and $\sim E^{-3}$ at high energies.  The transition energy is around $2 (m_e/m_\mu) E_\mu$, which is about 2 GeV for the muons in Super-K.  Again, reasonable agreement is seen.  Electrons are also produced in pair production, and this component is the same as the positron spectrum.

The gamma-ray spectrum is rather flat, which follows from the form of the bremsstrahlung differential cross section, which is $\sim 1/E$~\cite{pdg}.  Except at the highest energies, showers initiated by gamma rays are subdominant.

The rate of hadronic showers is small because muons primarily lose energy by electromagnetic processes.  The dominant hadrons made directly by muons are pions, with comparable numbers of each charge.

Relative to a mono-energetic muon spectrum,  using the full Super-K spectrum in Fig.~\ref{shower_spectrum} (as we do) leads to only modest differences.  At the highest energies, the differential cross sections for delta-ray production and pair production quickly increase with muon energy~\cite{pdg}.  Consequently, the electron and positron production are increased at high energies.  For the other particles and energies, the differences are less.

The spectra of muon daughter particles, and hence the showers they induce, favor low energies.  For electromagnetic showers, because of the dominant rate of delta-ray production, the total spectrum has a $\mathrm{d}N/\mathrm{d}\!\log_{10} E \sim 1/E$ shape.  The delta-ray spectrum does not stop at 0.1 GeV but keeps rising at lower energies.  These low-energy delta rays do not shower or make isotopes, but they do create an almost continuous light intensity on top of the flat light profile from the muon, with little variation between muons.  The hadronic shower spectrum is relatively flat, with a wide peak near 0.4 GeV.  (The hadronic component in electromagnetic showers is of comparable, but smaller frequency.)  Though hadronic showers are rare, with rate below 1\% of all showers, they are quite important for producing isotopes.  To obtain the expected number of all showers per muon above a given energy, we integrate the curves in Fig.~\ref{shower_spectrum}; above 0.1, 1, 10, and 100 GeV, we obtain 3.6, 0.4, 0.04, and 0.003.  For each muon, there will be Poisson fluctuations in the number of showers.  In Sec.~\ref{subsec:individualshowers}, we calculate the energy distributions of showers weighted by isotope and light production.

\section{Isotopes are born in showers}\label{sec:correlation}

In our previous paper~\cite{Li2014}, we showed that isotopes are typically not produced directly by muons, but rather by their low-energy secondaries.  (An exception, discussed in Sec.~\ref{subsec:energyloss}, is stopping $\mu^-$.)  The isotope yields follow from convolutions of secondary-particle path-length spectra with isotope-production cross sections.  Neutrons and pions are the most important secondaries for producing background isotopes --- those that decay with detectable signals in Super-K.  In contrast, gamma-ray secondaries primarily produce harmless isotopes --- those that are stable or decay invisibly.  We focus on background isotopes.  

In this section, we show that most isotopes are produced in rare, individual showers.  On one hand, this is not surprising, because isotope production increases with secondary particle path length, and showers produce many secondaries in a short distance.  On the other hand, it has been assumed that isotopes are made continuously along the muon tracks.

A consequence of our claim is that isotopes are produced at random but specific locations, coincident with showers, along muon tracks.  This picture is different from one where we average over muons (as in Ref.~\cite{Li2014}), so that isotopes are produced nearly uniformly along the muon track.  As we show, showers can identify and localize isotope production, because showers are detectable through their Cherenkov signals.

Using position information for preceding showers, the cuts to reduce spallation decays need to be applied only to a short section of the muon track that effectively covers the shower.  Compared to most Super-K analyses, where cuts are made along the whole muon track, this would allow decreased backgrounds without increased deadtime.  With the same deadtime, cutting less volume allows a longer time cut, improving background rejection.  A version of this technique was pioneered by Super-K in a search for the diffuse supernova neutrino background~\cite{Bays2012}, and it was shown to work to remove spallation backgrounds down to decay energies of 16 MeV.  Our goal, besides giving the first explanation of why this technique works, is to show how to extend it down to 6 MeV, where the spallation rate is much higher, and apply it to solar neutrino studies.

In the remainder of this section, we show how light and isotope production correlate with muon energy loss, how they causally depend on the initiating particle and energy of showers, and how well in principle these showers could be identified and localized.  We calculate the distributions of products --- showers, light, and isotopes --- from individual muons.  Super-K could use these distributions, following their likelihood approach, to assess the probability that an observed signal is of a particular origin, e.g., if a low-energy event is signal or background (and, if so, which muon was likely the cause).

\subsection{Muon energy loss leads to light and isotopes}\label{subsec:energyloss}

There can be several independent showers along a muon track.  When that is the case, detecting each shower and measuring its energy would require geometric reconstruction.  It is easier to measure the total visible muon energy loss through the total Cherenkov light intensity.  The true muon energy loss is slightly larger than the apparent energy loss because of the reduced light yield of hadronic showers.

Increased muon energy loss results in greater path length in secondaries and, hence, more Cherenkov light.  Most of the radiative energy loss goes into producing electromagnetic showers, and the subsequent electrons are contained in the detector.  Thus Super-K can measure the energy loss (but not the absolute energy) of a throughgoing muon by the total light deposited.  The radiative part can be obtained by subtracting the amount expected from a muon with the minimum energy loss (greater than the minimum ionization rate because these muons are relativistic).  Even in the rare cases where there are hadronic energy losses, the total light is a reasonably faithful (better than a factor of 2; see above) measurement of the muon energy loss.

Increased muon energy loss results in more isotopes, also due to more secondaries.  However, there is an important difference: {\it While light production is common, isotope production is rare}.  Most background isotopes in water are produced by low- to medium-energy hadronic secondaries (see Fig.~7 of Ref.~\cite{Li2014}), which are rarely produced and which are subdominant to electromagnetic secondaries.  Recall that hadronic showers always induce electromagnetic showers (but not vice versa), and that the light from the latter is typically dominant. 

\begin{figure}[t]
    \begin{center}                  
        \includegraphics[width=\columnwidth]{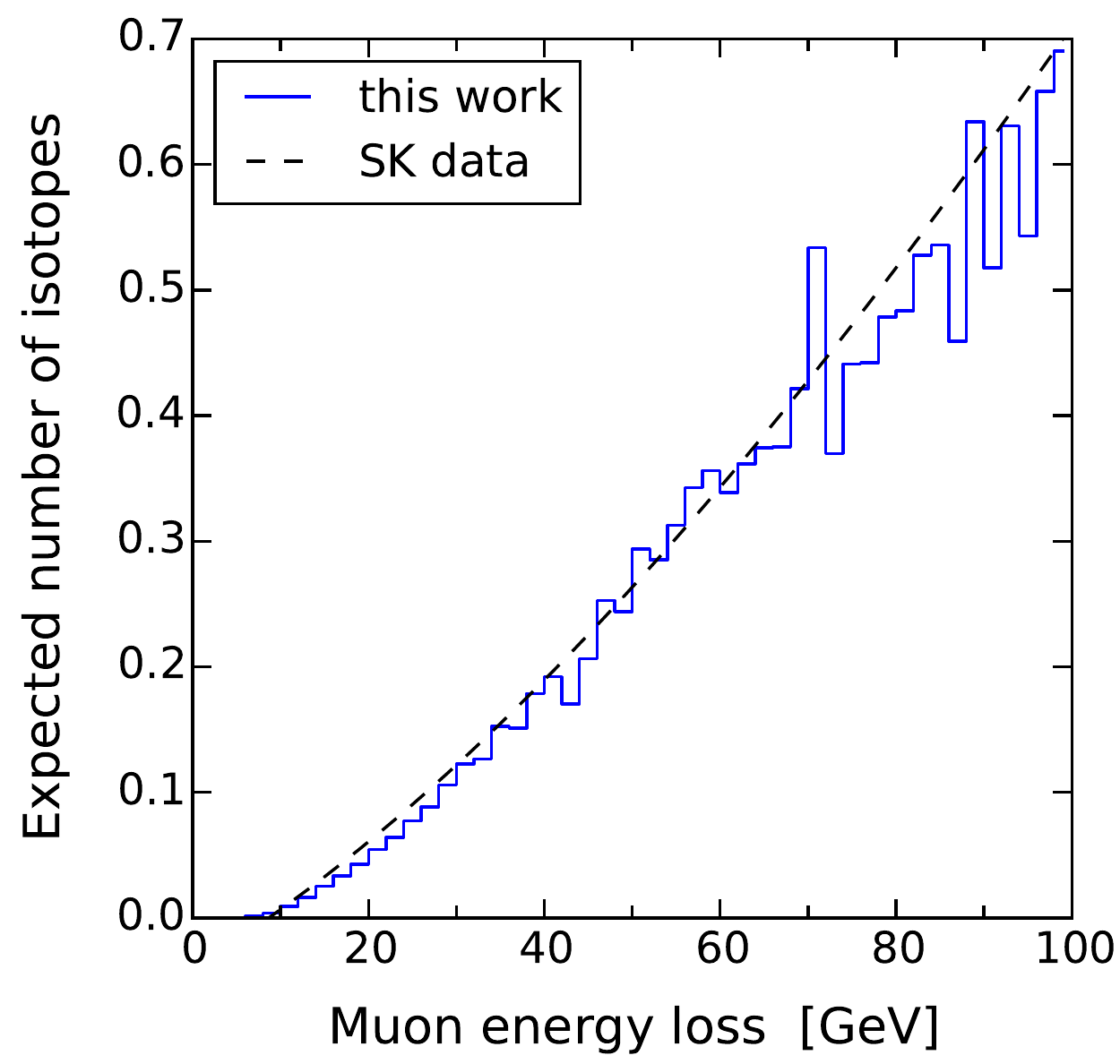}
        \caption{The expected number of background isotopes as a function of the total muon energy loss.  The solid line is our calculation assuming vertical throughgoing muons that travel 32.2 m in the FV, and the dashed line is the (corrected to match assumptions) Super-K measurement.}
        \label{light_iso}
    \end{center}
\end{figure}

Figure~\ref{light_iso} shows our calculation of how the production of background isotopes increases with total muon energy loss.  We also show the Super-K measurement, which is part of their likelihood function for spallation cuts, defined in terms of residual charge, $Q_\mathrm{res}$, the number of detected photoelectrons in excess of that expected from a muon with the minimum energy loss.  We made conversions between residual charge and energy loss for which we could find only an approximate factor (1000 photo-electrons $\simeq$ 130 MeV~\cite{Ishino}).  We assume that the Super-K results are for the expected number of isotopes per muon and that they need to be corrected by a factor $1/0.1$ because only a fraction of isotopes are included by the cuts used to select spallation events; Refs.~\cite{Hosaka2006,Koshio} are not clear about either point.  We obtain 0.1 by direct calculation, not {\it ad hoc} adjustment; this arises from two factors, each $\simeq$ 0.3, for a time cut of $\lesssim$ 0.1 s and an energy cut of $\gtrsim$ 7 MeV.  In addition, we assume that all muons are vertically throughgoing.  Nevertheless, our estimates should be reasonably accurate.  The good agreement with the Super-K measurement indicates that our simulation is correctly modeling muon energy loss and isotope production.  

This simple figure illustrates several important points that hint at the physics of isotope production in showers.  First, the average production rate of background isotopes is small, even for large muon energy losses.  (The yield of harmless isotopes is about ten times larger.)  Second, this function becomes nonzero only beyond about 7 GeV, which is where muon radiative loss processes start~\cite{Li2014}.  Third, the curve rises faster than linearly for low values of muon energy loss.  We separately checked individual isotopes, and found that they follow the same trend as the total shown in the figure.

\begin{figure}[t]
    \begin{center}                  
        \includegraphics[width=\columnwidth]{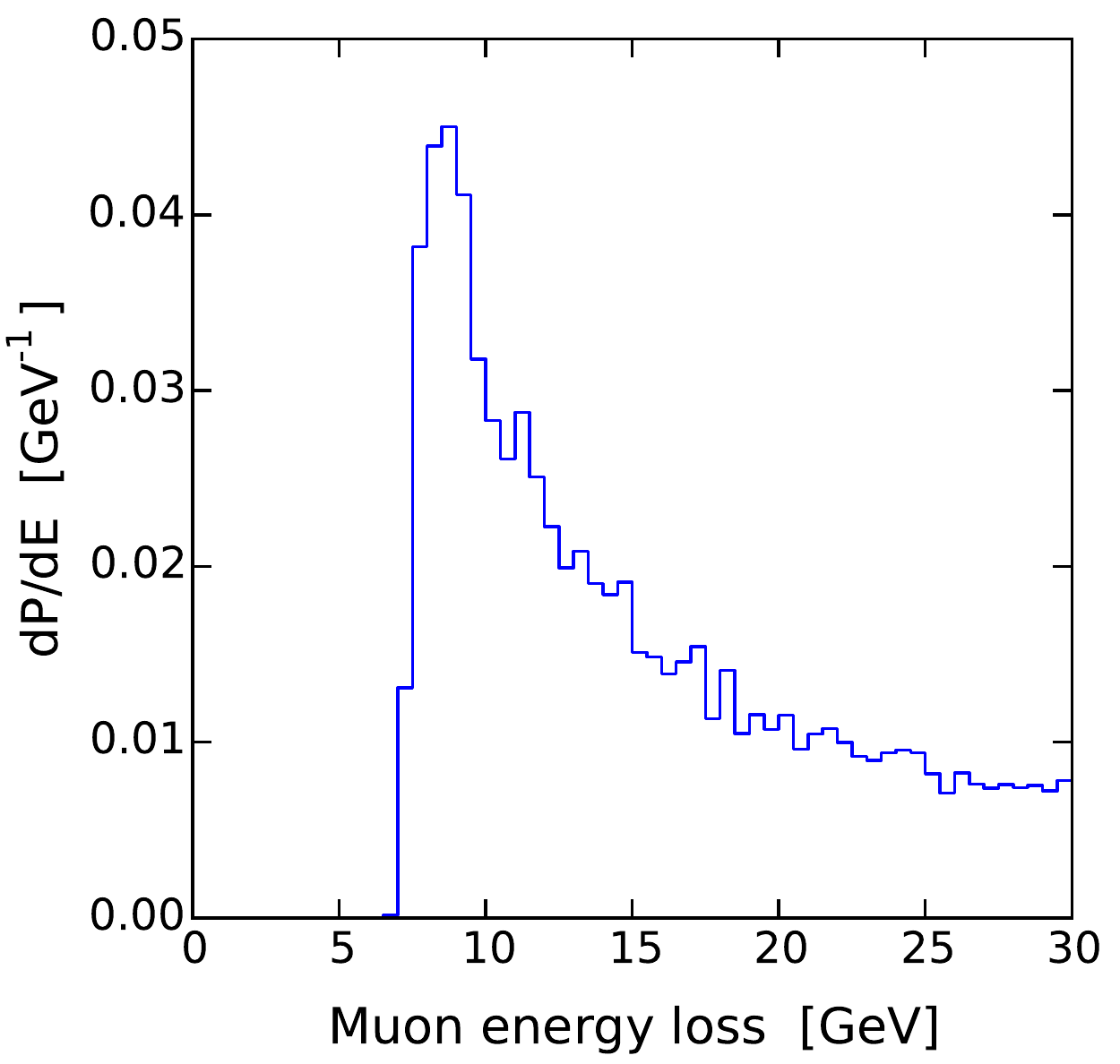}
        \caption{Probability distribution of muon energy losses, weighted by the isotope yield per muon.  The curve falls off slowly, extending to hundreds of GeV.  We show it up to 30 GeV, assuming that isotopes with large energy loss can be cut separately.}
        \label{isotope_muon_rate}
    \end{center}
\end{figure}

There are two possible shower frequency scenarios that could lead to Fig.~\ref{light_iso}.  A point common to both simply follows from Poisson statistics, which we illustrate using an energy loss of 30 GeV.  Because the number of background isotopes per muon is 0.1 on average, the number of isotopes produced is 1 for 1 muon and is 0 for 9 muons.  However, Fig.~\ref{light_iso} does not tell us the frequency of showers that make isotopes.  Small electromagnetic showers are more frequent and less efficient at making isotopes.  If the isotopes were made by such showers, then the number of showers per muon would be $\sim$ 1, with a fraction $\sim$ 0.1 of them making isotopes.  Hadronic showers or very energetic electromagnetic showers are less frequent and more efficient at making isotopes.  If the isotopes were made in these showers, then the number of such showers per muon would be $\sim$ 0.1 with a fraction $\sim$ 1 of them making isotopes.  Distinguishing these scenarios is important.  If isotope-producing showers were small in energy and common in position, then spallation cuts would have to be applied along the whole muon track; in contrast, if these are big and rare, they could be localized to short regions along the muon track.  The physics of isotope production by showers determines which shower energy range is most important.

Although isotope production rises with muon energy loss, the frequency of muon energy loss falls steeply (see Fig.~2 of Ref.~\cite{Li2014}).  When the muon energy loss is large, strong cuts can be applied without increasing deadtime because the frequency of such events is low.  For example, {\it muon energy losses of 30 GeV or more lead to $\simeq$ 60\% of the isotopes in Super-K, while being only 2\% of all muons}.  A simple cylinder cut along the muon track could thus eliminate a majority of isotopes with little deadtime.  Using a radius of 3 m and delay of 20 s for just the muons with large energy losses, Super-K could cut $\simeq$ 58\% of isotopes with only $\simeq$ 4\% deadtime.  (For comparison, a radius of 1 m and a delay of 20 s, applied to all muons, would cut $\simeq$ 80\% of isotopes with $\simeq$ 20\% deadtime, close to what Super-K achieves with more sophisticated likelihood techniques.)  In Ref.~\cite{Bays2012}, Super-K introduced a new cut on ``showering muons," defined to be those with an energy loss $\gtrsim 60$ GeV; for these, all data in the next 4 s from the whole detector are discarded.  We estimate that this has substantially worse efficiency and deadtime than our proposed new cut.

Our investigations also demonstrate that no spallation cuts are necessary along the tracks of stopping muons.  Muons with low energy ($\lesssim$ 7 GeV) lose all their energy by ionization in the FV.  Because their energies are low, they do not typically lose energy by radiative processes.  Consequently, very few isotopes (0.4\% of all isotopes) are produced along their tracks.  At the ends of their tracks, however, negative muons can capture on oxygen, which can lead to nuclear breakup.  Thus, a separate cut for stopping muons where only events inside a sphere centered on the end of the muon track are rejected would be highly efficient with minimal deadtime (Super-K has such a cut for $^{16}$N~\cite{Blaufuss2001}).

Figure~\ref{isotope_muon_rate} shows a histogram of muon energy loss for muons that make isotopes, weighted by the number of isotopes produced.  The shape of the histogram is the frequency of muon energy loss in Super-K (Fig.~2 of Ref.~\cite{Li2014}) multiplied with the yield of isotopes from muons in Super-K (Fig.~\ref{light_iso}).  We focus on a small energy range (below 30 GeV), assuming that high-energy-loss muons can be cut as suggested above.  This figure shows that the most probable energy loss for isotope production is small.  However, there is a long tail, extending to hundreds of GeV.  Once the energy loss range is constrained to a reasonable range, the cut should be optimized for small energy losses.

\subsection{Individual showers are the cause}\label{subsec:individualshowers}

When we average over muons and along their tracks, as above, light and isotope production are correlated through the total muon energy loss.  Here we break that energy loss into individual showers, and detail how light and isotopes are causally related to showers with different injection energies and initiating particles.  These relationships determine the geometry of the spallation cuts. 

\begin{figure}[t]
    \begin{center}                  
        \includegraphics[width=\columnwidth]{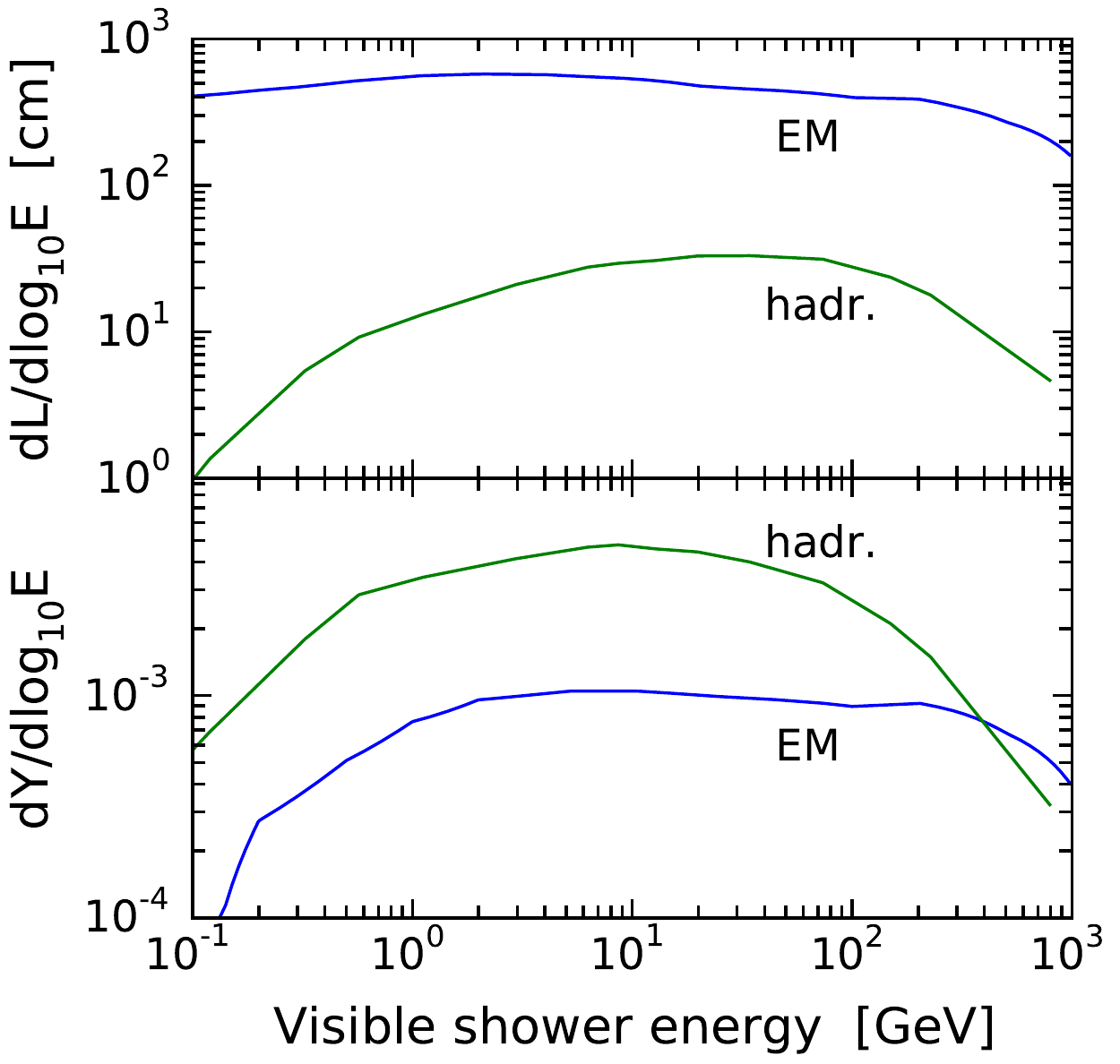}
        \caption{Light yield (top panel) and background isotope yield (bottom panel) for showers of different types and energies.  The ``EM'' curves include showers initiated by $e^{\pm}$, $\gamma$, and $\pi^0$; the ``hadr.'' curves include showers initiated by charged pions, kaons, and other hadrons.  Yields are per vertical throughgoing muon in Super-K, taking into account the cosmic-ray muon spectrum.}
        \label{efficiency}
    \end{center}
\end{figure}

Figure~\ref{efficiency} shows our results for the average yields of light and isotopes made by showers as a function of energy.  To calculate how muon-induced showers produce light and isotopes, we obtain the number spectra of daughter particles produced directly by muons using Fig.~\ref{shower_spectrum}, then multiply these number spectra with the yields of light and isotopes by showers with those energies.  This approach accounts for nearly all the daughter particles from the radiative energy losses of muons; we discuss the exceptions below.  We define showers initiated by $\pi^\pm$ (including a small contribution from kaons and other hadrons) to be hadronic, and those initiated by $e^\pm$, $\gamma$, or $\pi^0$ to be electromagnetic.  To compare to experiment, we use visible energy, determined from the total Cherenkov light (proportional to the integrated path length above the Cherenkov thresholds) made by relativistic particles (see Sec.~\ref{subsec:em},~\ref{subsec:hadronic}).  At injection energies below 0.1 GeV, the curves drop off because showers do not form; at energies above $10^3$ GeV, they drop off because such injection energies are rare.

An immediate conclusion is that {\it light production is strongly dominated by electromagnetic showers}, which are by far the most common.  Another is that {\it background isotope production is somewhat dominated by hadronic showers}, even though they are much more rare.  

The light yield distributions depend on the physics of muon energy loss and of shower development.  At lowest order, the light yield $\mathrm{d}L/\mathrm{d}\!\log_{10}E$ follows $E\mathrm{d}N/\mathrm{d}\!\log_{10}E$, which can be obtained by multiplying Fig.~\ref{shower_spectrum} by $E$.  Electromagnetic showers in this energy range are primarily induced by delta rays from muons, and their frequency falls as $\simeq 1 \,(\mathrm{GeV}\!/E_0)$ shower per energy decade per muon traveling the length of the Super-K FV (3220 cm).  The light yield of an electromagnetic shower rises as $\simeq 500 {\rm\ cm\ } (E_0/ \hspace{0.05em}\mathrm{GeV})$.  In combination, the result is $\simeq 500$ cm, almost independent of shower energy.  (This continues to even lower energies, dropping slightly, due to low-energy delta rays.)  That is, 5000 cm of light is equally likely to be from one 10 GeV shower or ten 1 GeV showers; these cases can be distinguished by reconstruction of the light profile along the muon track.  Hadronic showers in this energy range are primarily induced by pions from muons; the rate relative to delta-ray production is $\sim 10^{-2}$ near 1 GeV but increases steeply with injection energy.  Hadronic showers convert most of their energy to electromagnetic showers, which produce nearly all of the light, and this efficiency increases with injection energy.  The light yield for hadronic showers as a class is therefore quite suppressed and is not as flat as for electromagnetic showers.  At low energies, this variation is especially pronounced because of low pion production by muons.  The total light yields (integrated over energy) provide an important check of our calculation.  The average light yield per muon is $\simeq 2000$ cm, corresponding to a radiative energy loss of about 4 GeV, or a total energy loss of about 11 GeV, in good agreement with the average we found in Ref.~\cite{Li2014}.

The isotope yield distributions depend on similar physics, plus the interaction cross sections of secondaries with nuclei.  Although the frequency of hadronic showers is low, the neutrons and pions they produce are quite efficient at making background isotopes.  (Above a total muon energy loss of about 30 GeV, this efficiency is so high that it becomes possible that 2 or more isotopes are produced in the same shower, which would allow their clear identification and localization as background events.)  EM showers make isotopes mostly through the neutrons and pions they produce, but also directly through gamma rays.  The shapes of the isotope distributions are similar to each other and to the light distributions, but there are some important differences.  Low-energy hadronic showers are especially efficient (per injected energy) at making isotopes, because they convert less of their energy to electromagnetic showers; low-energy electromagnetic showers are especially inefficient because of the threshold energy needed to induce hadronic showers.  The shape of the isotope production curve here is closely related to that in Fig.~\ref{isotope_muon_rate}.  Here we consider the energy of individual showers, each of which contributes to the radiative energy loss; the total energy loss in Fig.~\ref{isotope_muon_rate} includes about 7 GeV for ionization energy loss.  Also, here we use a $\log$ axis, which stretches out small radiative losses, and a log derivative, which has the effect of multiplying the shape by a factor $\sim E$.

These facts show why the total muon energy loss and isotope production are correlated but not causally connected.  Most of the detected muon energy loss comes from electromagnetic showers.  In contrast, most isotopes are made by hadronic showers.  Both types of shower increase with muon energy loss.  The correlation between energy loss and isotope production is not simply linear because of the steep rise of isotope production as a function of shower energy at low energies.  Even at the level of individual showers, the production of light and isotope production are not completely causal.  The isotope production per shower is typically low, which means the presence of a shower does not necessarily indicate the production of an isotope.  However, when an isotope is produced, it is almost always preceded by light from a shower, and that is what makes the Super-K background-reduction technique possible.

How well the Super-K technique works depends on the frequency of showers that make isotopes.  {\it The drop in isotope production in low energy showers shown in Fig.~\ref{efficiency} is crucial.}  Few isotopes are made by low-energy showers, which are common, or low-energy delta rays, which are near continuous.  From Fig.~\ref{shower_spectrum}, we calculate that the integrated rate of showers becomes $\simeq$ 1 per muon when the minimum daughter particle energy is $\simeq$ 0.4 GeV.  Because almost all isotopes are made by higher-energy showers, this technique can work with minimal confusion about which shower to associate with an isotope.  If low-energy showers had produced too large a fraction of isotopes, the associated showers would be too frequent along the muon track for this technique to be practical.

\begin{figure}[t]
    \begin{center}                  
        \includegraphics[width=\columnwidth]{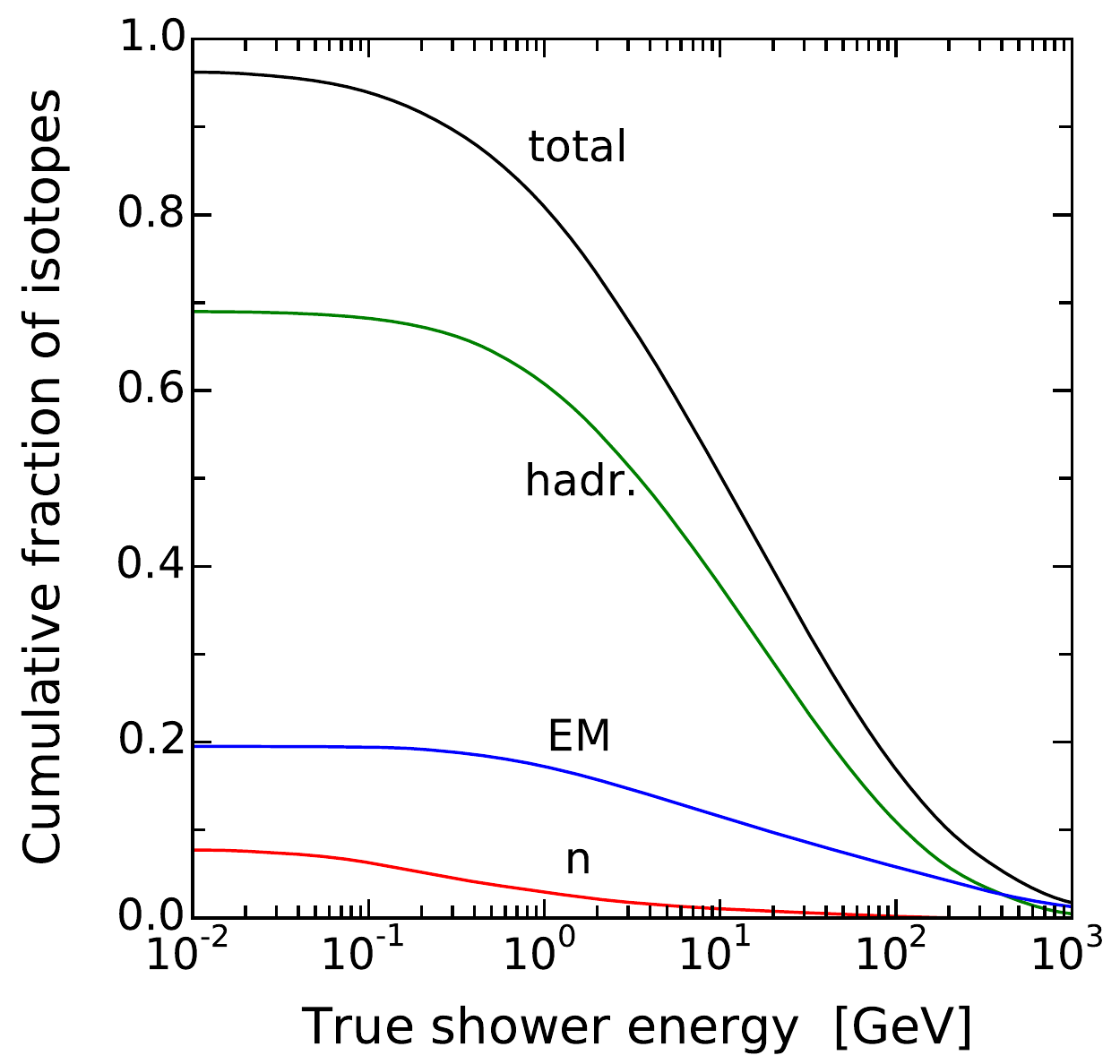}
        \caption{Cumulative fraction of background isotopes produced by showers above a given injection (daughter particle) energy.  All curves are normalized by the number of isotopes produced per muon, excluding those produced directly by muons.  The hadronic and electromagnetic components are as in Fig.~\ref{efficiency}, and the neutron component is discussed in the text.  ``Total'' means the sum of all processes.}
        \label{efficiency_cumulative}
    \end{center}
\end{figure}

How well the Super-K technique works also depends on the fraction of background isotopes produced in showers.  Figure~\ref{efficiency_cumulative} shows the fraction of isotopes contained in showers above a given energy.  The curves are integrations of the isotope yield curves in Fig.~\ref{efficiency}, now using true shower energy.  The hadronic and electromagnetic components shown in Fig.~\ref{efficiency_cumulative} are the same as in Fig.~\ref{efficiency}.  The neutron component, not shown in Fig.~\ref{efficiency} because it produces so little light, is special.  Above a few hundred MeV, neutron secondaries act as part of the hadronic component.  At lower energies, they can induce ``neutronic'' showers, where neutrons collide with nuclei, ejecting neutrons (and protons), continuing the process, producing isotopes but very little light; this accounts for only a few percent of isotopes.

{\it Figure~\ref{efficiency_cumulative} shows that nearly all isotopes are made in showers induced by muon daughter particles}, which is also crucial for this technique.  (We exclude isotopes made directly by primary muons, which make 3\% of all isotopes, mostly through processes that then produce identifiable showers.)  Above 0.01 GeV, we recover 96\% of the isotopes that are not directly produced by primary muons.  Within the precision of our calculations, this agrees well with the isotope yield in Ref.~\cite{Li2014}, where we did not separate the processes leading to isotope production.  This supports our claim that nearly all isotopes are made in showers.  In future work, we will show that nearly all of the showers in Fig.~\ref{efficiency_cumulative} are identifiable.

\subsection{Showers can tag isotope production}\label{subsec:tagshower}

The results above show that isotopes are almost always produced in showers, and that these showers are detectable by their light.  The probability of isotope production increases with shower energy, though it is small at the most important energies.  These facts agree with the usual Super-K spallation likelihood function, for which isotope production increases with the total muon energy loss.  If this energy loss can be localized to a shower, it will allow the cut to be applied to a shorter section of muon track.  The success of the Super-K cut technique depends on the fraction of isotopes produced in identifiable showers.

\begin{figure}[t]
    \begin{center}                  
        \includegraphics[width=\columnwidth]{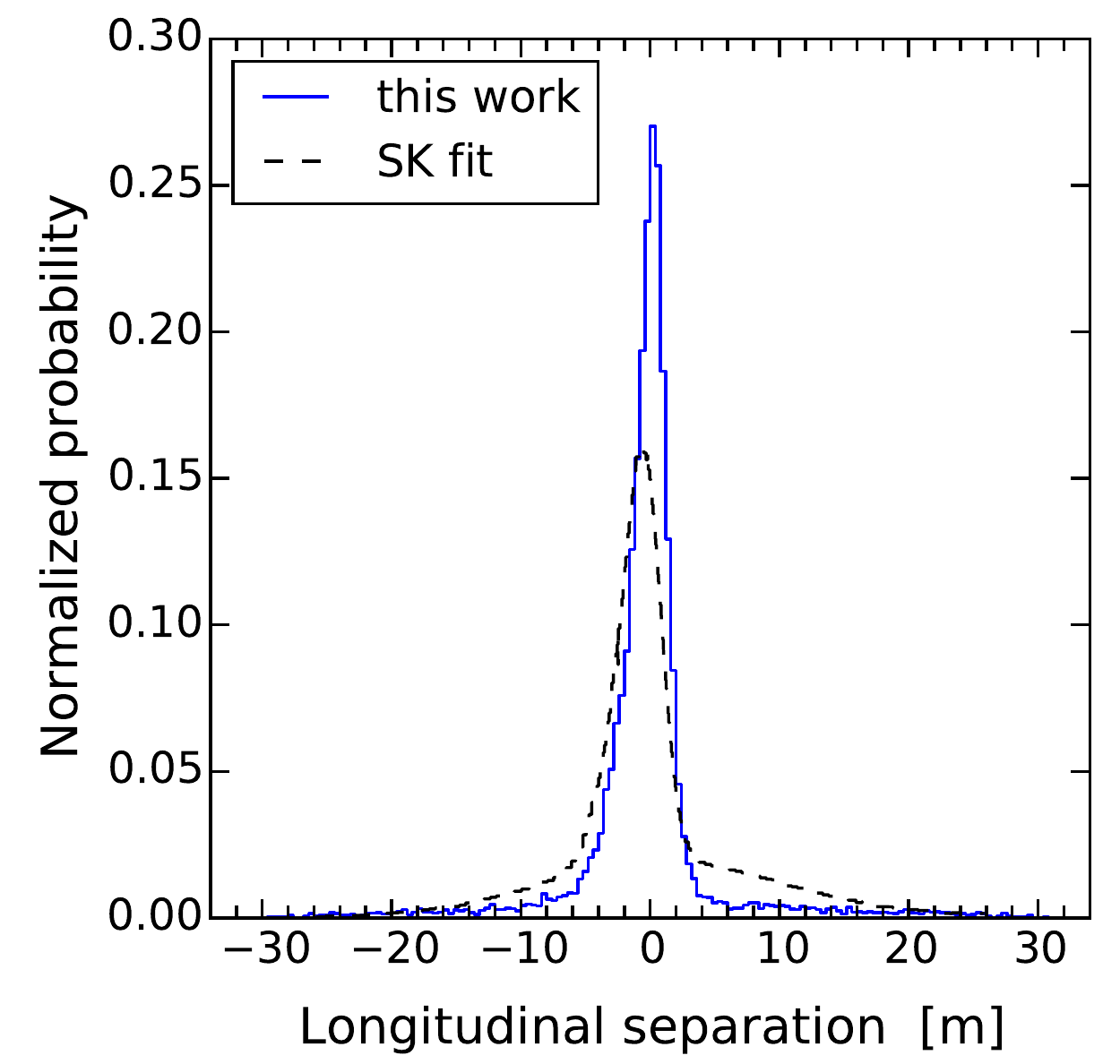}
        \caption{The longitudinal separation distribution between showers and isotopes.  The solid line is our calculation assuming a perfect shower reconstruction technique, and the dashed line is the Super-K measurement.  The Super-K technique already works very well but could be significantly improved.}
        \label{seperation_distribution}
    \end{center}
\end{figure}

Figure~\ref{seperation_distribution} shows the distribution of separation distances between the peak of the muon light profile and isotope production point.  We first describe our calculation in detail, and then compare to the Super-K result.  For each individual muon, the peak of the muon light profile is taken to be the point of the maximum charged particle path length along this muon track.  We define the $z$ coordinate to increase along the muon track, beginning at the top of the detector, and the separation distance to be the $z$ position of the shower minus that of the isotope.  For calculating the maximum light position, we use a binning of 50 cm, comparable to the position resolution in Super-K at low energies; other reasonable choices give similar results.  When more than one isotope is produced, we compute the separation distances for each.  Because of how the distribution is defined and would be used in a likelihood approach, there is no conceptual problem with having more than one isotope produced by one muon.  Practically speaking, the most common such scenario should be two isotopes produced in a rare, high-energy shower. 

The separation distribution has a large peak and small tails.  {\it The peak comes from the case where the isotope is produced in the largest shower along the muon track.}  The isotope production profile generally follows the shower longitudinal profile.  The peak in Fig.~\ref{seperation_distribution} is thus centered at zero.  The full width of the peak of $\simeq 4$ m at half-maximum follows from that of the longitudinal shower profiles; it extends further to the left because showers are longer after the peak than before.  Because the peak is quite sharp, it can define a new spallation likelihood function with stronger cuts over a shorter section of muon track, as empirically discovered in Ref.~\cite{Bays2012}.  The tails, which can be barely seen at separations of tens of meters, arise from cases where the isotopes and showers are uncorrelated.  As the distributions of showers and isotopes are nearly flat along the muon track, the tails have a well-defined shape --- a symmetric triangle peaked at zero separation.  For our calculation, we find that the area in this triangle is 13\% of the total.

To improve background rejection, it is important to understand the reasons for this uncorrelated component.  We find that $\simeq 3\%$ is due to isotope production accompanied by very little light; the parent particles are high-energy muons or low-energy neutrons made by them, in a ratio of about 1 to 2.  The largest portion, $\simeq 10\%$, is due to cases where the isotope is produced in a visible shower, but where there is a larger shower elsewhere on the muon track; we determine this by examining isotope-shower pairs with large separations.  As a check, we find that this portion increases if we increase the height of the simulated detector.   A key issue for reducing the uncorrelated component will thus be improving the identification of multiple independent showers along the muon track.  The uncorrelated component is as small as it is, even with this simple approach, because the expected number of showers per muon is small.

Figure~\ref{seperation_distribution} also shows the Super-K result from Ref.~\cite{Bays2012}.  Although it is similar, there are some important differences.  The most important is that the area in the tails is $\simeq 25\%$ instead of 13\%.  This excess is due to cases where the isotope is produced in a shower that would have been visible in our simulation but was not visible in the Super-K analysis, at least after the smearing effects of imperfect Cherenkov reconstruction.  We can approximately recover the Super-K fraction of $\simeq 25\%$ if we assume that showers below $\simeq 10$ GeV (muon energy losses below $\simeq 17$ GeV) cannot be reconstructed.  In addition, the peak and tails are not symmetric, which we think is due to problems with shower reconstruction, as discussed in our next paper.  Our estimates about the Super-K results are crude, as their analysis has low statistics and large bin widths (this could be improved by their using spallation decay energies lower than 16 MeV); the functions used to fit their data seem nonideal; the noted asymmetries cause uncertainties; and there is the possibility of differences in the selection of single-throughgoing muons in the Super-K analysis and in our simulations.

There are two major steps Super-K can take to strengthen the correlation between showers and isotopes.  First, they could attempt to reconstruct showers of lower energy.  A $\sim 10$ GeV shower more than doubles the light from a muon track, and we expect that much smaller showers could be identified.  If they can do this down to very low energies, their measured result should match what we obtained in our simulated data, and they could reduce $\simeq 25\%$ to $\simeq 13\%$.  Second, they could attempt to recognize multiple showers per muon, defining cut regions around each.  Because showers are relatively rare, it would probably be enough to reconstruct up to two showers.  If this were successful, they could reduce $\simeq 13\%$ down to $\simeq 3\%$.

In future work, we will show that it should be possible for Super-K to improve their reconstruction technique well enough to match our results in Fig.~\ref{seperation_distribution}, and then even further, i.e., reducing the tails of the distribution function with new methods.  This will allow significantly better background rejection.

\section{Conclusions and Future Work}\label{sec:conclusion}

Low-energy neutrino detectors could continue to provide invaluable information about the Sun, supernovae, and neutrino properties.  Prominent goals include the $hep$ solar flux, the DSNB flux, and the solar day-night mixing effect.  Super-K is large enough, but progress depends on reducing detector backgrounds.  In the energy range 6--18 MeV, the dominant background is from the beta decays of unstable nuclei produced by cosmic-ray muons and their secondaries.  Super-K has strong cuts to reduce these backgrounds, but the residual rates are large.

We are undertaking a multipart project to provide tools to significantly reduce these spallation backgrounds in Super-K.  Our project, based on a foundation of careful simulation and theoretical insights, is the most extensive such effort undertaken for any detector.  With modest adjustments, our results will be useful for other water-based detectors, e.g., WATCHMAN~\cite{Askins2015} and Hyper-Kamiokande~\cite{Abe2011hk}.  Since these detectors are likely to be shallower than Super-K, spallation backgrounds will be even more severe.  More generally, our results will provide valuable insights about backgrounds in other underground detectors for neutrinos, dark matter, and other rare processes such as neutrinoless double beta decay.

In our previous paper~\cite{Li2014}, we presented the first theoretical calculation of the spallation background yields in Super-K.  We focused on the steady-state background rates, averaged over muons and along their tracks.  We found that almost all isotopes are produced by secondary particles, and not the primary muons themselves.  Our predictions for the spallation decay backgrounds agree with Super-K aggregate data to within a factor of 2, which is very good and could be improved.  Our results provide new information about components, correlations, and production mechanisms that can be used to develop cuts that are more powerful than those based on empirical studies. 

Our next steps were inspired by a recent Super-K DSNB analysis~\cite{Bays2012}, where the Cherenkov light profiles associated with individual muons were measured.  These were found to vary along the muon tracks, showing peaks, with the positions of the peaks correlated with the sites of isotope production.  A new cut was developed using this correlation, and was shown to be effective for improving the DSNB search.  However, the cause for the variation in the light profile and its correlation with isotope production remained mysteries.  This new cut has not yet been used for solar neutrino analysis.  It seems very promising for reducing backgrounds without increasing deadtime.

In the present paper, we consider how isotope production varies between muons and along their tracks.  We break the process of muons producing isotopes into muons producing energetic daughter particles, these daughter particles inducing electromagnetic and hadronic showers, and these showers producing isotopes.  We provide details about each step and combine them in the end.  Our calculations here break our previous calculations~\cite{Li2014} into more steps, but agree in overall approach and results.

Our fundamental result is that showers are the key to explaining the correlation between muon light profiles and spallation backgrounds, as well as their total yields of spallation products in Super-K.  Showers produce electrons, which make Cherenkov light, and neutrons and pions, which make background isotopes.  In Fig.~\ref{efficiency}, we show how showers of different types and energies contribute to the production of light and isotopes.  Because of the high rate of electromagnetic showers, and the high efficiency of hadronic showers for making isotopes, electromagnetic showers strongly dominate light production and hadronic showers somewhat dominate isotope production.  Isotopes are nearly always proceeded by showers, though only a small fraction of showers produce isotopes.  With these results, we reproduce Super-K results on muon energy loss (Fig.~\ref{light_iso}), isotope production (Fig.~\ref{efficiency_cumulative}), and their correlations (Fig.~\ref{seperation_distribution}).  

We are the first to show that the background isotopes in Super-K are dominantly made in discrete, identifiable showers.  (It has long been known that isotope production is associated with muons with high radiative energy loss, e.g., Refs.~\cite{Nakahata1988,Hirata1989} and much subsequent work, but it had not been shown that these showers are rare enough and energetic enough to be identifiable, and that they account for the production of nearly all isotopes.)  Though this paper focuses on Super-K, our results have much more general applicability.

The calculations and insights of this paper and of Ref.~\cite{Li2014} can be used to define new cuts that should be very effective for solar and DSNB analyses.  Some could be implemented easily (the muon energy loss and stopping muon cuts in Sec.~\ref{subsec:energyloss}); others improve the technique of Ref.~\cite{Bays2012} (the efficiency of the technique depends on how well Super-K reconstructs the muon light profile, Sec.~\ref{subsec:tagshower}); and others need new development (our forthcoming papers).

We will soon demonstrate new ways to better identify showers.  As mentioned above, the Super-K reconstructed light profiles are inconsistent with what we expect from showers.  In our next paper, we identify the reason for this inconsistency and will demonstrate better ways to reconstruct muon Cherenkov light profiles.  In the Super-K reconstruction equation, which solves for the emission position of each individual photomultiplier hit, there are two possible solutions for the light from deflected electrons; we will show how to select the better solution, and that doing so improves the resolution.  In addition, we will show how to isolate shower light from muon light, which also helps significantly.  In subsequent papers, we will discuss new signals that can identify showers with even higher efficiency, followed by quantitative studies of the effects of new cuts on background rates and the implications for solar and supernova neutrino analyses.

\section{Acknowledgments}

S.W.L. and J.F.B. were supported by NSF Grants PHY-1101216 and PHY-1404311 to J.F.B.  We are grateful to Patrick Allison, Kirk Bays, Mauricio Bustamante, Anton Empl, Ciriyam Jayaprakash, Paolo Lipari, Masayuki Nakahata, Kenny Ng, Eric Speckhard, Michael Smy, Yasuo Takeuchi, and Mark Vagins for helpful discussions.

\end{document}